\documentclass[letterpaper, twocolumn, superscriptaddress,  amsmath, amssymb,longbibliography]{revtex4-1}
\usepackage[pdftex]{graphicx}
\usepackage{caption}
\usepackage{subcaption}
\usepackage{dcolumn}
\usepackage{bm}
\usepackage{color}
\usepackage{hyperref}
\usepackage{adjustbox}
\usepackage{color}
\usepackage{float}
\usepackage{framed}
\usepackage{esvect}
\usepackage{amsfonts,amssymb,amsmath,hyperref,hypcap,enumerate}
\usepackage{wasysym}
\usepackage{slashed}
\usepackage{tikz}
\usepackage{amsmath,amssymb,bm,graphicx,dsfont}
\usepackage[latin9]{inputenc}
\usepackage{amsmath}
\usepackage{slashed}
\usepackage{dsfont}
\usepackage{amstext}
\usepackage{amssymb}
\usepackage{amsbsy}
\usepackage{amsthm}
\usepackage{epsfig}
\usepackage{tikz}
\usepackage{bbm}
\usepackage{hyperref}
\usepackage{color}
\usepackage{multirow}
\usepackage{ulem}
\newcommand{\be}{\begin{equation}}
\newcommand{\ba}{\begin{align}}
\newcommand{\ee}{\end{equation}}
\newcommand{\bea}{\begin{eqnarray}}
\newcommand{\eea}{\end{eqnarray}}
\newcommand{\beq}{\begin{equation}}
\newcommand{\eeq}{\end{equation}}
\newcommand{\beqn}{\begin{eqnarray}}
\newcommand{\eeqn}{\end{eqnarray}}

\newcommand{\la}{\langle}
\newcommand{\ra}{\rangle}

\newcommand{\ket}{\rangle}
\newcommand{\bra}{\langle}

\renewcommand{\vec}[1]{{\bf #1}}
\renewcommand{\hat}[1]{{\widehat #1}}

\def\nn{\nonumber\\}

\begin{document}
\title{
Electric polarization as a nonquantized topological response and boundary Luttinger theorem}
\author{Xue-Yang Song}
\affiliation{Department of Physics,  Harvard University,
Cambridge, MA 02138, USA}
\affiliation{Perimeter Institute for Theoretical Physics, Waterloo, ON N2L 2Y5, Canada}
\author{Yin-Chen He}
\affiliation{Perimeter Institute for Theoretical Physics, Waterloo, ON N2L 2Y5, Canada}
%\affiliation{Department of Physics,  Harvard University, Cambridge, MA 02138, USA}
\author{Ashvin Vishwanath}
\affiliation{Department of Physics,  Harvard University,
Cambridge, MA 02138, USA}
\author{Chong Wang}
\affiliation{Perimeter Institute for Theoretical Physics, Waterloo, ON N2L 2Y5, Canada}
%\affiliation{Department of Physics,  Harvard University, Cambridge, MA 02138, USA}
\date{\today}

\begin{abstract}

We develop a nonperturbative 
%topological 
approach to the bulk polarization of crystalline electric insulators in $d\geq1$ dimensions. 
Formally, we define polarization via the response to background fluxes of both charge and lattice translation symmetries.
 In this approach, the bulk polarization is related to properties of magnetic monopoles under translation symmetries. Specifically, in $2d$ the monopole is a source of  $2\pi$-flux, and the polarization is determined by the crystal momentum of the $2\pi$-flux. In $3d$ the polarization is determined by the projective representation of translation symmetries on Dirac monopoles. Our approach also leads to a concrete  scheme to calculate polarization in $2d$, which in principle can be applied even to strongly interacting systems. For open boundary condition, the bulk polarization leads to an altered   `boundary' Luttinger theorem (constraining the Fermi surface of surface states) and also to modified  Lieb-Schultz-Mattis theorems on the boundary, which we derive. 
%and is more generally related to Lieb-Schultz-Mattis type of quantum anomalies for the boundary low-energy theory.

\end{abstract}

\date{\today}

\maketitle

%\tableofcontents

\section{Introduction}

The bulk electric polarization of an insulator is a concept of fundamental importance in condensed matter physics. Polarization as a bulk quantity in ferroelectricity, piezoelectricity (polarization induced by mechanical stress) etc, have been widely studied in traditional solid state physics.
An important theoretical discovery that related (the change in) polarization to a geometric Berry phase\cite{vanderbiltbook, Resta_rmp, RestaVanderbilt, Martin72,Kallin84,Vanderbilt,Resta95,Ortiz96,Martin97,Martin98} reveals its profound connection to topology properties in quantum systems such as quantum Hall effects and topological insulators.  The precise definition and interpretation of polarization, however, is a subtle issue (see, for example, Refs.~\cite{RestaVanderbilt,WatanabeOshikawa}). 

Intuitively the polarization measures the density of electric dipole moment in the bulk. Polarization density in a $d-$dimensional crystalline system with volume $V$ and charge density $\rho(r)$ reads,
\begin{equation}
\label{def}
\hat P= \frac{1}{V}\int d^d r \rho(r)\vec r.
\end{equation}
However a surface charge distribution will induce a non-vanishing change in polarization per volume, due to the position operator in the definition. Powerful methods that avoid such an issue have been established to calculate the polarization, at least within the independent electron approximation of band theory\cite{vanderbiltbook,Resta_rmp, Martin97,Resta92, Martin98,vanderbilt_1993}. Periodic boundary condition was instead adopted and polarization  was calculated using Wannier functions of the occupied bands. It is also demonstrated that for a generic insulating system with interactions, the change in polarization during an adiabatic evolution is well-defined and given by the integrated bulk currents, which could be further expressed as a many-body Berry phase \cite{OrtizMartin}.

In this work we develop a more topological approach to define and to \textit{directly} measure the bulk polarization in arbitrary spatial dimensions. Specifically we use a topological term to define polarization (per unit cell volume) in any dimensions, and explore its consequences for both the periodic and open boundary conditions. For the periodic boundary condition, the polarization determines the properties of \textit{magnetic monopoles} under translation symmetries (such as momenta). For the open boundary condition, the bulk polarization determines the degree of \textit{Luttinger theorem violation} on the boundary, and more generally is related to quantum anomalies of the boundary low-energy theory.  Our approach applies to short-range entangled systems of interacting fermions and bosons (or spins) as long as there are lattice translation symmetries and a conserved $U(1)$ charge with a charge gap (i.e. an insulator).

The paper is organized as follows. In Sec.~\ref{sec1d} we motivate a $1$d expression of the topological term for polarization and discuss its various implications and issues for the generalization to higher dimensions. Sec.~\ref{sectopterm} introduces the notion of translation gauge fields and the topological term for polarization as a central result of this work. Sec.~\ref{secmonopole} explores the connections of translation properties of magnetic monopoles and polarization density in $2$d and $3$d, implicated by the topological term in Sec~\ref{sectopterm}. Sec.~\ref{secboundary} discusses open systems where bulk polarization modifies the boundary Luttinger theorem. The relation of surface charge distribution to bulk polarization is further clarified through this discussion. Lieb-Schultz-Mattis (LSM) type constraints descending from our topological term on topologically ordered systems and defect scenarios are also discussed. Sec.~\ref{secconclusion} summarizes the results and provides physical arguments underlying the entire work. The appendices contain details on calculation recipes, subtleties on polarization, derivation and discussion of anomalies and numerical results.

Before discussing the main results,  we note that there is a feature in the definition of polarization when the unit cell has a nontrivial geometric structure. The polarization $\vec P$ consists of two parts: a classical electric dipole moment within each unit cell and an extra part that measures inter-unit-cell entanglement  \cite{Kudin2,Bardarson_2016} -- the latter is denoted as $\tilde{\vec{P}}$ in Ref.~\cite{WatanabeOshikawa}. Both $\vec{P}$ and $\tilde{\vec{P}}$ have been discussed in the literature depending on context. We review these notions briefly in Appendix~\ref{PPtilde}.  %dependent on unit cell choice), which describes the intra- and inter- unit cell charge distribution, respectively. 
Our results below can be applied to both the full $\vec P$ and to $\tilde{\vec{ P}}$ as long as we adopt the  appropriate calculation scheme as  described below.

\section{Polarization in $1$d}
\label{sec1d}

Polarization has been \cite{Resta, resta_98,aligia_99,OrtizMartin} formulated in $1$d systems of size $L$ at integer fillings in terms of the expectation value of the large gauge transformation operator on the ground state (GS), i.e.
\begin{equation}
2\pi P=\textrm{Im ln}\langle GS| e^{i2\pi \hat P}|GS\rangle,
\end{equation}
where $\hat P$ is the dipole moment density in Eq.~\eqref{def} and the operator induces a gauge transform on the electron operator $c_r\rightarrow c_r e^{i2\pi r/L}$. Watanabe and Oshikawa \cite{WatanabeOshikawa} showed its equivalence to the Berry phase for a flux piercing process of a $1$d ring where flux $\theta$ adiabatically increases from $0$ to $2\pi$ (under an appropriate gauge choice of the ground state).  
The electromagnetic field $A_x(r)=\theta/L$ increases as the flux pierces the system. In the $1+1$D action, the $2\pi P$ phase accumulation  associated with the time-dependent $A_x(t)$ can be naturally written as $P \int dxdt \partial_tA_x$, where $\int dx A_x(t)$ increases from $0$ to $2\pi$. Hence the polarization can be considered as an electromagnetic response  defined by a topological $\Theta$-term in the low energy (IR) (setting $\hbar=c=e=1$):
\be
\label{1dtheta}
S^{1D}_{P}=P\int dxdt\,(\partial_tA_x-\partial_xA_t),
\ee
where we added another term $\partial_x A_t$ to keep the gauge invariance. The flux-piercing process induces a change of $2\pi$ in $\int dxA_x$  and hence a phase of $2\pi P$ in the action.
This term can also be motivated by the fact that a dipole moment $\vec{d}$ couples to electric field as $-\vec{d}\cdot\vec{E}$. Since $\int dA=\int dxdt\,(\partial_tA_x-\partial_xA_t)$ is always an integer (the first Chern number) multiple of $2\pi$ on a closed $2$-manifold, $P$ is defined mod $1$. The periodicity of $P$ can be understood on the lattice by noticing that shifting an integer charge by one lattice unit (we set to be $a=1$) in every unit cell is equivalent to a relabeling of lattice coordinates and should not have any physical effect.

The polarization, defined via Eq.~\eqref{1dtheta}, has several consequences. First for periodic systems, as discussed above, an adiabatic flux-threading process where $\int dx\, A_x$ changes by $2\pi$ %\footnote{The flux threading  should be followed by a large gauge transform that brings the Hamiltonian back to the original form.} 
 leads to a Berry phase $\Phi=2\pi P$ from the space-time path integral of Eq.~\eqref{1dtheta}. This Berry phase is in principle a measurable quantity and is sometimes used as the definition of polarization in one dimension\cite{OrtizMartin,Resta}. For open boundary condition Eq.~\eqref{1dtheta} becomes a boundary term $\pm P\int dt A_t$, which represents a fractional charge $q=\pm P$ (mod $1$) at each boundary -- the mod $1$ condition comes from the fact that one can always deposit an integer charge on the boundary without affecting the bulk. This is consistent with the intuitive connection between polarization and dipole moment.
 
We note that for the Berry phase to be well-defined, the ground state is required to return to itself up to a phase after an adiabatic flux threading, i.e. the ground state space should be non-degenerate. As a counter example, for fractional filling cases, according to Lieb-Schultz-Mattis theorem, a gapped ground state must break translation invariance. For a rational filling $p/q$ (irreducible fraction), it could be remedied by threading $2\pi q$ flux and measure the polarization as the Berry phase divided by $q$, modulo $1/q$\cite{aligia_99}.

Raising this to higher dimensions poses some challenges. A simple generalization of the electromagnetic response term Eq.~\eqref{1dtheta} to higher dimensions does not produce a topological term. One can consider it as a Berry phase term, and measure the polarization through the Berry phase of a flux-threading process (say in the $x$-direction) similar to that in $1D$. The Berry phase, however, is given by
\be
\label{BerryExtensive}
\Phi_x=\frac{V}{L_x}2\pi P_x \hspace{10pt} ({\rm{mod}}\hspace{2pt} 2\pi),
\ee
where $V=L_xL_y...$ is the system volume and $L_x$ is the length in $x$-direction. If we assume \textit{lattice translation symmetries} (which we do for the rest of the paper), the intensive quantity $\vec{P}$ can be extracted from the $L$-dependence of $\Phi$ (but simply dividing by $V/L_x$ will not work since the phase is defined mod $2\pi$). However it raises the conceptual question whether $\vec{P}$ itself bears any physical meaning. For example, for a two-dimensional crystalline insulator with $L_y=2N$ ($N\to \infty$ in thermodynamic limit) and a polarization density $P_x=1/2$, the Berry phase from Eq.~\eqref{BerryExtensive} is always trivial. Is there a formula for the  polarization  in this case? One can always define polarization by starting with a reference state with a known polarization (for example where polarization is constrained by symmetries) and connect it to the Hamiltonian of interest by an adiabatic path, and integrating the currents obtained while connecting the initial and final states. However this algorithm requires defining such an adiabatic path and is conceptually different from a direct measure of  polarization that we seek.

A similar issue appears with open boundary condition: the density of dipole moment, which is the classical definition of polarization, is given by the surface charge density. Unlike the $1D$ case where the boundary charge is robustly determined mod $1$, the surface charge density in higher dimensions can be continuously tuned by boundary perturbations (for example a boundary chemical potential). It then appears that the boundary does not necessarily reflect the bulk polarization. An exception was observed in Ref.~\cite{vanderbilt_1993}: when the boundary is gapped and non-degenerate, the boundary charge density faithfully represents the bulk polarization mod $1$.

As we shall see, a topological approach is needed because polarization cannot be measured by local probes -- something global, such as symmetry fluxes or physical boundaries, has to be introduced. This justifies the use of the term ``topological response", even though the response itself (the polarization) is in general not quantized and hence its value will change in response to  symmetric perturbations. The familiar electromagnetic  polarizability (the $\Theta$-term) in $3d$ also falls into this category when the time-reversal and mirror symmetry (or more generally symmetries that invert an odd number of spacetime coordinates) are absent.

%%%%%%%%\section{Polarization from a Topological term}

\section{Polarization from topological terms} 
\label{sectopterm}
Interesting IR physics can often be probed by the response to background gauge fields. In the study of polarization density in $d>1$, the relevant symmetries include charge conservation and lattice translation symmetries, so we shall consider coupling the system to gauge fields associated with these symmetries. For charge conservation the gauge field is simply the electromagnetic field $A_{\mu}$. For each translation symmetry $\mathbb{Z}$, say in the $i$'th direction, we introduce a $\mathbb{Z}$-gauge field $x_i$. This ``translation gauge field"\cite{ThorngrenElse} is less familiar so we review below. The gauge field $x_i$ is locally flat ($dx_i=0$) so only its Wilson loops $\int_{C_1} x_i\in \mathbb{Z}$ over loops  (or $1$-cycles $C_1$) in space-time is meaningful -- formally this means that $x_i\in H^1(M,\mathbb{Z})$ where $M$ is the space-time manifold. Furthermore, just like the Wilson loops in other gauge theories, the integer $\int_{C_1}x_i$ measures the number of $\hat{x}_i$-translations one has to go through to travel across $C_1$. To be more concrete consider a path integral description, with dynamical degrees of freedom $\psi$ (bosonic or fermionic) defined in continuous time $t\in[0,T)$ and on discrete lattice sites $s$ in space:
\be
e^{-iS_{eff}[A,x_i]}=\int D[\psi(s,t)]\exp{\left(-i\sum_s\int dt \mathcal{L}_s[\psi,A]\right)},
\ee
where we have used locality and translation symmetries to write the Lagrangian as a sum of local terms of identical form, $\mathcal{L}_s[\psi,A]$, which involves only fields near site $s$. We take periodic boundary conditions in space and time (so $M$ is a torus). The translation gauge fields enter the partition function by specifying exactly how the periodic boundary contitions are taken: 
\bea
\psi(s,t)&=&\psi\left(s+\hat{x}_j\int_ix_j,t\right); \nn
\psi(s,t)&=&\psi\left(s+\hat{x}_j\int_tx_j,t+T\right).
\eea
We now explain these equations in more details. The Wilson loop of $x_i$ in the $\hat{x}_i$ direction gives the lattice size $\int_ix_i=L_i$. For $j\neq i$ the number $\int_ix_j$ measures how much the slice of the lattice at $x_i=L_i$ is displaced along the $\hat{x}_j$ direction before it is identified with the slice at $x_i=0$. Similarly the time component $\int_tx_i$ measures the displacement of the entire lattice at $t=T$ before identified with $t=0$. In other words, while the ``longitudinal" parts of the translation gauge fields measure the lattice size, the ``transverse" parts measure the quantized shear strains of the lattice in both space and time. We can also consider a $(d-2)$ dimensional defect in space, around which $\int x_i=n\neq0$: this is simply a lattice dislocation with Burgers vector $\vec{B}=n\hat{x}_i$.

The translation gauge field $x_i$ is closely related to the concept of tetrad in the theory of elasticity\cite{DZYALOSHINSKII198067}, which has been used to characterize three dimensional integer quantum Hall effect recently\cite{Tetrads1,Tetrads2, Tetrads}  and torsions in Weyl semimetals \cite{torsion_huang}. Consider embedding the lattice into a continuous space, so that each site $s$ can be assigned a continuous coordinate $\vec{u}_s$. We can treat $\vec{u}$ as a field, then the tetrad $\vec{\nabla}u_i$ will have all the properties of $x_i$ discussed above, and can be used as a representation (a gauge choice) of $x_i$. The gauge invariant properties of $x_i$ such as the Wilson lines, however, do not depend on how the lattice is embedded into a continuous space. In this sense the $x_i$ gauge field measures the topological part of elasticity response. Another straightforward way is to consider the strain tensor $\partial_i d_j(\vec r)$ in elasticity theory\cite{kleinert, Manjunath_2020}, where the displacement fields $\vec d(\vec r)$ for site $\vec r$ are defined modulo the lattice spacing, i.e. a relabeling of sites by an integer vector $\vec N(\vec r)$ does not make a physical difference. This invariance calls for the gauge field $x_i$,
\begin{align}
    d_i(\vec r)&\rightarrow  d_i(\vec r)+ N_i(\vec r),(N_i\in \mathbb Z)\nonumber\\
    \nabla  d_i&\rightarrow \nabla d_i+x_i,
\end{align}
where $x_i=\nabla N_i$ is defined on a discrete lattice and can be viewed as the translation gauge field.

Now recall that the electric polarization in $1d$ can be defined through the topological term $P\int dA$ (Eq.~\eqref{1dtheta}). The natural generalization to higher ($d+1$) dimensions is the following term:
\be
\label{gaugeresponse}
S_{Polar}=\sum_i(-1)^{i+1}P_i\int x_1\wedge... x_{i-1}\wedge dA \wedge x_{i+1}...\wedge x_d.
\ee
Here $\wedge$ should really mean cup product $\smile$ for discrete cohomology instead of the usual wedge product, but the distinction does not matter for our purpose. 

We now give some justifications for Eq.~\eqref{gaugeresponse} as a definition of bulk polarization. {{First, it is the only topological term involving $dA$ and $x_i$ that is first order in the field strength $dA$, as we expect for the polarization.}} Each component of polarization $P_i$ is defined mod $1$ since the integral always gives integral multiples of $2\pi$ on closed manifolds (the term is therefore a topological $\Theta$-term), and is in agreement with the intuition that shifting integer charges by one lattice unit does not have physical effect. We note that for a system with spin degeneracy, in principle two EM fields $A_{\uparrow,\downarrow}$ could be used to couple to the phases of spin up, down electron operators, respectively. Consequently, two topological terms with coefficients $P_{\uparrow,\downarrow}$ are present and each of the two polarization quantities is defined mod $1$. Total polarization density $P=P_{\uparrow}+P_{\downarrow}$ is defined mod $2$.  When evaluated for a uniform electric field $\vec{E}$ on a perfect lattice (free of dislocation and shear strain) of size $L_1\times L_2...$, this term becomes
\be
S_{Polar}(L_i)=(V/L_i)\sum_iP_i\int dtdx_i E_i,
\ee
which agrees with the expectation that the total polarization is $(V/L_i)P_i$ when the system is viewed as $1d$ in $\hat{x}_i$ direction. In addition, if $\vec{P}$ has a time-dependence $\vec{P}(t)$, then by taking derivative with respect to $\vec{A}$ from the above action we obtain the charge current $\vec{j}=\partial\vec{P}/\partial t$, which agrees with physical expectations and is sometimes used as a practical way to define polarization.  If $\vec P$ is spatially dependent, say varying in $\hat x_i$ direction, taking derivative with respect to $A_t$ on the term $(-1)^{i+1} \int P_i x_1\wedge\cdots x_{i-1}\wedge dA \wedge x_{i+1}\cdots\wedge x_d$ gives $-\rho=\partial_{xi} P_i$, agreeing with the relation $-\rho=\nabla\cdot\vec P$.\footnote{This integrating by part may leave a boundary term which accounts for the change of surface bound charge $\sigma_i$ due to spatial variation of polarization through $\sigma_i=\vec P\cdot \hat n$, where $\hat n$ is the normal vector of the surface.} We emphasize that while the special or temporal variations of $\vec{P}$ results in locally measurable quantities like $\rho$ or $\vec{j}$, the more subtle constant piece of $\vec{P}$ comes with an intrinsically topological nature and needs to be defined via the topological term Eq.~\eqref{gaugeresponse}.

We note that the translation gauge field can also be used for magnetic translations. Suppose we have a lattice system in which each unit cell traps a $U(1)$ magnetic flux $\phi$ in the $xy$-plane, then in our formulation the Dirac quantization condition for the $U(1)$ gauge field is now changed to 
\be
\int_{C_2}(dA-\phi x\wedge y)=0\hspace{5pt}({\rm{mod}}\hspace{2pt}2\pi),
\ee
where $C_2$ represents arbitrary $2$-cycles in spacetime. Most of our discussions in this paper will be equally applicable for magnetic translations as long as the above modified Dirac quantization condition is imposed.

%Topological response terms like Eq.~\eqref{gaugeresponse} appear frequently in the study of symmetry-protected topological (SPT) phases\cite{KapustinCobordism}. However in our case $P_i$ takes continuous value and is therefore not protected, which is not surprising since polarization can change adiabatically by tuning the Hamiltonian.

%%%%%%%%\section{Monopole properties from polarization}

\section{Polarization and monopoles} 
\label{secmonopole}
We now examine the consequences of the polarization as defined through Eq.~\eqref{gaugeresponse} on a closed manifold (like periodic boundary condition). Motivated by the $1d$ case, it is useful to consider instantons of the $A$ field. In $1d$ the instanton is the familiar adiabatic flux-threading, a smooth configuration in spacetime. In higher dimensions the instantons become operators supported on $(d-2)$ dimensional sub-manifolds in space, with $\int dA=2\pi$ on the two complementary spatial dimensions. For $d=2$ this is simply a unit flux insertion in space, and for $d=3$ it corresponds to a unit flux tube in space whose open ends become Dirac monopoles. On the $(d-2)+1$ manifold of the instanton, the topological terms reduces to the following Dijkgraaf-Witten\cite{DijkgraafWitten} type:
\be
\label{InstantonAction}
S_{in}=\sum_i(-1)^{i+1}2\pi P_i\int x_1\wedge...x_{i-1}\wedge x_{i+1}...\wedge x_d.
\ee

Let us look at some physically relevant examples. At $d=2$ we obtain
\be
S_{in,2d}=2\pi\int dt (P_1x_2^{(t)}-P_2x_1^{(t)}),
\ee
which means that the $2d$ monopole -- a point operator in space -- carries ``charge" of $2\pi (-P_2,P_1)$ under translation symmetries in $\hat{x}_{1}$ and $\hat{x}_2$, respectively. But ``charge" under translation symmetry is simply the crystal momentum. We then conclude that in $2d$ the monopole carries lattice momentum 
\be
\label{MMomentum}
\vec{k}_{\mathcal{M}}=2\pi(-P_2,P_1)=2\pi\hat{z}\times\vec{P}.
\ee 
It may be helpful to have some simple semi-classical picture here. Consider a $2\pi$-flux quanta spread uniformly over a region  much larger than the lattice unit. We can then consider the momentum of such a monopole $\vec{k}_{\mathcal{M}}$, i.e. the Berry phase from moving the unit flux configuration by one lattice unit. Equivalently we can consider the many-body momentum of the fermions $\vec{k}_e$ under the flux configuration, which would be the inverse of the monopole momentum. Now imagine a semi-classical continuum system, a non-uniform electric field is induced during the turning-on of the magnetic flux, which then induces a momentum on a small electric dipole moment $\vec{d}$ according to $\delta\vec{k}\sim\int dt (\vec{d}\cdot\nabla)\vec{E}\sim -\int dt \vec{d}\times(\nabla\times\vec{E})\sim\vec{d}\times\vec{B}$. Since the dipole density is given by $\vec{P}$, we have
\be
\label{classical}
\vec{k}_e=-\vec{k}_{\mathcal{M}}=\int d^2r\vec{P}\times \vec{B}=2\pi\vec{P}\times\hat{z},
\ee
which is what we obtained from the topological term.

We can use the monopole momentum as a practical way to calculate electric polarization in $2d$. We outline the calculation scheme here and discuss more details in Appendix~\ref{2DNumerics}. Consider a $2d$ crystalline insulator on a torus, and smoothly spread a total magnetic flux of $2\pi$ on the lattice -- say $2\pi/L_xL_y$ flux per plaquette. The total momentum of the many-electron system $\vec{k}_e$ can be measured from the ground state wave function, as Berry phase factors associated with lattice translations, and from Eq.~\eqref{classical} we have $\vec{k}_e=2\pi\vec{P}\times\hat{z}$. The virtue of this calculational scheme is that it is well defined (although possibly complicated) even for strongly correlated systems, where band theory techniques cannot be used.

\begin{figure}
 \captionsetup{justification=raggedright}
    \centering
     \adjustbox{trim={.18\width} {.3\height} {.2\width} {.27\height},clip}
   { \includegraphics[width=0.7\textwidth]{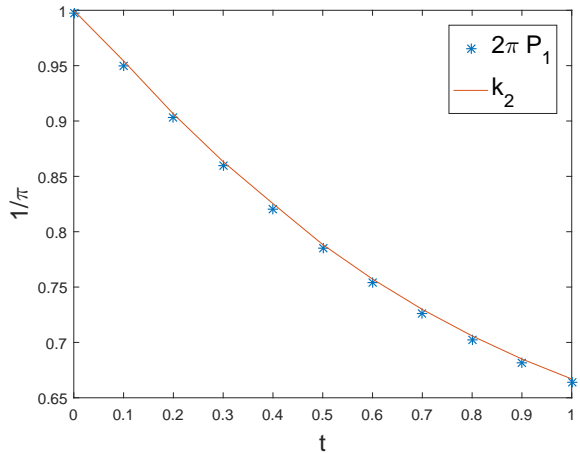}}
    \caption{The polarization $2\pi P_1$ calculated from band theory and monopole momentum $k_2$ along orthogonal direction always agree as one tunes a parameter $t$. Details can be found in Appendix~\ref{2DNumerics}. }
    \label{fig:numerict}
\end{figure}

Another consequence of the connection between polarization and monopole momentum is that we can now define polarization in $2d$ even in the presence of gapless Dirac fermions. The only subtlety is that with a unit flux in space, each Dirac cone contributes a zero-energy mode, leading to multiple degenerate ground states depending on which zero modes are occupied. For each of the degenerate monopoles we can nevertheless define its lattice momentum and interpret it as a bulk polarization, which then also depends on the zero mode fillings. In Fig.~\ref{fig:numerict} we report a numerical calculation of the polarization using the monopole momentum, for a lattice system of gapless Dirac fermions with a specific choice of zero mode filling. The same polarization can also be calculated using the standard method from band theory which we also report in Fig.~\ref{fig:numerict}. The two results clearly agree as we vary a parameter $t$ in the Hamiltonian. More details of the calculational recipe and the lattice model can be found in Appendix~\ref{2DNumerics}. In fact, there is a long history of numerically calculating monopole momenta for lattice Dirac fermions, motivated by the study of monopole operators in Dirac spin liquids\cite{alicea_2008,ranvishwanathlee,hermele_2008,monopole2}. We showed that what this calculation really produces is the polarization of the underlying Dirac fermions.

\begin{figure}
 \captionsetup{justification=raggedright}
    \centering
     \adjustbox{trim={.2\width} {.28\height} {.1\width} {.13\height},clip}
   { \includegraphics[width=0.8\textwidth]{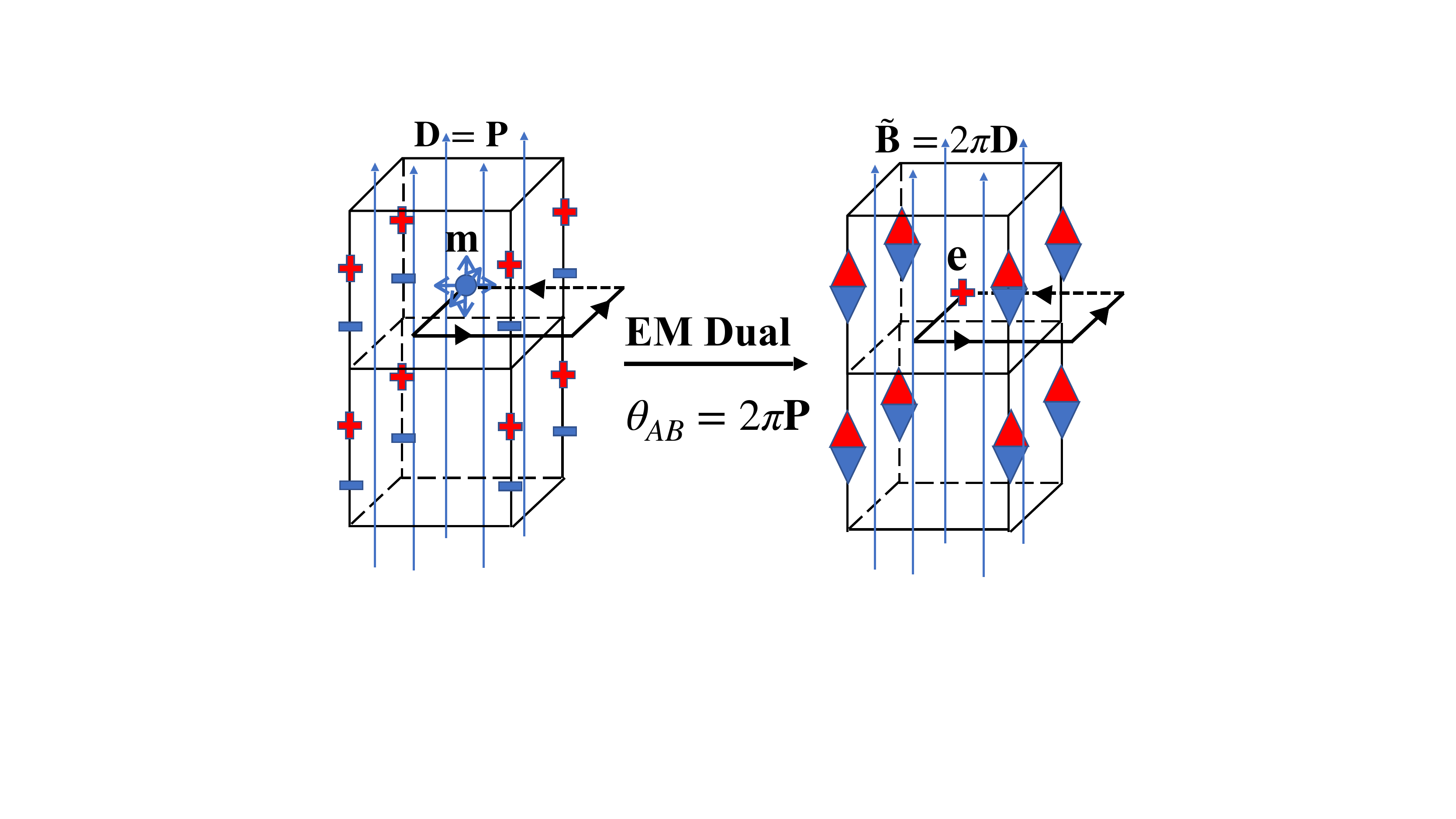}}
    \caption{The semi-classical picture relates projective representation of monopoles to polarization by electro-magnetic duality that exchanges magnetic monopole and electric charge.  The electric displacement field (left) $\vec D=\vec P$ is mapped to magnetic field (right) $\tilde {\vec B}=2\pi \vec D$ and the monopole to an electric charge. The Aharonom-Bohm (AB) phase $\theta_{AB}$ seen by the electric charge is proportional to magnetic field(right), hence the monopole Berry phase proportional to displacement field(left).}
    \label{fig:3d_em}
\end{figure}

For $d=3$ the term in Eq.~\eqref{InstantonAction} becomes a two-dimensional integral
\be
S_{in,3d}=\pi\epsilon^{ijk}\int P_ix_j\wedge x_k,
\ee
which describes a $2\pi$ flux loop decorated with a $1d$ topological phase enriched by translation symmetries. This leads to nontrivial boundary modes when the flux loop has open ends, which are nothing but Dirac monopoles. The boundary mode is characterized by a projective representation\cite{Pollmann2010,Chen2011,Schuch2011} of translation symmetries, namely translations in different directions commute up to a phase when acting on a Dirac monopole: 
\be
\label{3dMonopole}
T^{-1}_jT^{-1}_iT_jT_i={\rm{exp}}(i2\pi\epsilon^{ijk}P_k).
\ee
 This also has a simple semi-classical picture shown in Fig.~\ref{fig:3d_em}. Consider a $3d$ continuum system with polarization density $\vec{P}$. The polarization leads to an electric displacement field $\vec{D}= \vec{P}$. A magnetic monopole sees the $\vec{D}$ field as an effective ``dual magnetic field" $\tilde{\vec{B}}=2\pi\vec{D}=2\pi\vec{P}$. A particle moving in an effective magnetic field realizes translation symmetries projectively, namely different translation operations commute up to a phase factor according to Eq.~\eqref{3dMonopole}.

Similar to the $2d$ case, the relation Eq.~\eqref{3dMonopole} is relevant for $U(1)$ quantum spin liquids in three dimensions, described at low energy by an emergent Maxwell $U(1)$ gauge theory that is potentially realized, for example, in quantum spin ice materials\cite{savary_2017}. Our results indicate that the monopoles in a $U(1)$ spin liquid will carry projective translation quantum numbers if the emergent electric charges form an insulator with nontrivial polarization density.

It was known from earlier approaches that polarization is related to other topological quantities including Hall conductance and magneto-polarizability (the axion $\Theta$-angle). In Appendix~\ref{HallTheta} we show that these connections can be understood easily using the polarization-monopole connection.

We summarize the connection between bulk polarization and monopole (instanton) properties in $d=1,2,3$ in Table.~\ref{Summary}.

\begin{table}[htp]
\begin{center}
\begin{tabular}{|c|c|c|c|}
\hline
  & Monopole property & Polarization  \\ \hline
1D & Berry phase & $\Phi=2\pi P$  \\ \hline
2D & Momentum & $\vec{k}_{\mathcal{M}}=2\pi\hat{z}\times\vec{P}$  \\ \hline
3D & Projective momentum & $T^{-1}_jT^{-1}_iT_jT_i={\rm{exp}}(i2\pi\epsilon^{ijk}P_k)$  \\ \hline
\end{tabular}
\end{center}
\caption{ Polarization density $\vec{P}$ is related to the properties of the monopoles in dimensions $d=1,2,3$.  }
\label{Summary}
\end{table}

%%%%%%%%%\section{Boundary Luttinger theorem}

\section{Boundary Luttinger theorem and anomaly}   
\label{secboundary}
We now explore the consequences of bulk polarization for open boundaries.  Consider a boundary at $x_1=0$ separating the vacuum at $x_1>0$ and the polarized bulk at $x_1<0$, which preserves all translation symmetries except the one along $\hat{x}_1$. The $\Theta$-term in Eq.~\eqref{gaugeresponse} becomes a boundary term
\be
\label{polarizedcharge}
S_{\partial}=-P_1\int_{\partial}A\wedge x_2...\wedge x_d.
\ee
The meaning of this term can be seen by taking functional derivative with $A_0$: it simply means a (fractional) charge density of $\rho_{\partial}=P_1$ on the boundary. If the boundary has trivial dynamics in the IR, namely with a unique gapped ground state, then Eq.~\eqref{polarizedcharge} is the only nontrivial term in the IR description of the boundary. This is the well known statement that for a trivially insulating boundary, the charge density is given by the bulk polarization density mod $1$.\cite{vanderbilt_1993}

In general, depending on details at the boundary, the boundary can also host nontrivial low energy degrees of freedom. Let us first consider the simplest scenario: a Fermi liquid metal on the boundary. In this case the boundary charge density $\rho$ is obviously not fixed by bulk polarization $P$ (in direction perpendicular to the boundary) since it can be continuously tuned by perturbations that live only on the boundary. But we expect the Fermi surface volume $V_F$ to be tuned simultaneously with the charge density following $\Delta V_F/(2\pi)^{d-1}=\Delta\rho$ from Luttinger theorem. We therefore expect
\be
\label{LuttingerViolation}
\rho=\frac{V_F}{(2\pi)^{d-1}}+\vec P \cdot \vec{\hat n}\hspace{5pt}({\rm mod}\hspace{2pt} 1),
\ee
where $\vec{\hat n}$ is the normal vector of the boundary. From this relation the polarization density $P$ can be viewed as a source of Luttinger theorem violation on the boundary. Alternatively, we can view $V_F$ as a ``quantum correction" to the classical expectation of $P=\rho$. 

The ``boundary Luttinger theorem" Eq.~\eqref{LuttingerViolation} can be understood using an anomaly-matching argument. It is useful to first phrase the usual Luttinger theorem in terms of anomaly matching, following ideas similar to those in Ref.~\cite{Oshikawa}. Consider a theory in $(d-1)$ space dimensions of low-energy fermions near a Fermi surface, where fermion modes far away from the Fermi surface have been integrated out already. We couple the background gauge fields $A$ and $x_i$ minimally to these fermions and denote the action as $S_{FS}[\psi,A,x_i]$. It is known that under a large gauge transform, in which the real space Wilson loop along the $\hat{x}_i$ direction $\int_{C_i} A$ changes by $2\pi$, the total crystal momentum of these low energy fermions changes by\cite{Oshikawa} 
\be
\label{momentumchange}
\Delta \vec{K}=2\pi\frac{V_F}{(2\pi)^{d-1}}L_1...L_{i-1}\hat{x}_iL_{i+1}...L_{d-1}.
\ee
This means that the theory $S_{FS}[\psi,A,x_i]$ is not invariant under large gauge transforms. Instead, under a gauge transform $A\to A+d\alpha$, the low energy theory near the Fermi surface transforms as
\be
\label{Anomaly}
S_{FS}[\psi,A,x_i]\to S_{FS}[\psi,A,x_i]+\frac{V_F}{(2\pi)^{d-1}}\int d\alpha \wedge\prod_ix_i.
\ee
To see Eq.~\eqref{momentumchange} from Eq.~\eqref{Anomaly}, simply recall that $\int_{C_i}x_i=L_i$ and that the total momentum along $\hat{x}_i$ is the coefficient of $\int dt x_i$. Eq.~\eqref{Anomaly} is also related to the familiar chiral anomaly in $(1+1)$ dimension, which we briefly explain in Appendix~\ref{FSAnomaly}. Now for a purely $(d-1)$ dimensional system that is not the boundary of another space, we should add a background term
\be
\label{SFull}
S_{Full}=S_{FS}[\psi,A,x_i]-\frac{V_F}{(2\pi)^{d-1}}\int A\wedge\prod_ix_i,
\ee
so that the full theory is gauge invariant. The meaning of the counter term, as we discussed under Eq.~\eqref{polarizedcharge}, is simply a charge density of $\rho=V_F/(2\pi)^{d-1}$ -- this is nothing but the familiar Luttinger theorem!

It is now straightforward to extend to the case of Eq.~\eqref{LuttingerViolation}. Consider a Fermi liquid on the $(d-1)$ dimensional boundary of a $d$ dimensional bulk, with Fermi volume $V_F$ and boundary charge density $\rho$. The ``surface" theory reads
\bea
S_{Full}&=&S_{FS}[\psi,A,x_i]-\rho\int A\wedge\prod_ix_i, \nn
&=&\left\{S_{FS}[\psi,A,x_i]-\frac{V_F}{(2\pi)^{d-1}}\int A\wedge\prod_ix_i\right\} \nn
&&-\left(\rho-\frac{V_F}{(2\pi)^{d-1}}\right)\int A\wedge\prod_ix_i.
\eea
The collection in $\{...\}$ is gauge invariant, but the last term is not if $(\rho-V_F/(2\pi)^{d-1})\notin\mathbb{Z}$. We should therefore view the last term as a polarization term in $d$ space dimensions, hence Eq.~\eqref{LuttingerViolation}. The lesson is that bulk polarization does not directly give a boundary charge density, rather it leads to a boundary quantum anomaly. In Appendix \ref{square_numeric} we numerically study a free fermion model on square lattice and verify that Eq.~\eqref{LuttingerViolation} is always satisfied across a range of parameters with qualitatively different edge behaviors. Eq. \eqref{LuttingerViolation} also applies with $\vec{P}$ replaced by $\vec{\tilde P}$ if we also replace the  bound charge density by the excess charge density.

When viewed as an anomaly-matching condition, Eq.~\eqref{LuttingerViolation} can also be applied to surface states other than Fermi liquids -- we simply need to replace $V_F$ by the appropriate anomaly indicators of the low energy theories. Namely we demand $\rho=n_A+P$ mod $1$ where $n_A$ is the anomaly indicator of the low energy effective theory. For example, for rational values of $\rho-P$ the anomaly can be matched by a gapped ground state with intrinsic topological order, which typically hosts nontrivial quasiparticles with fractional electric charge. In such states the anomaly is encoded in how the topologically nontrivial excitations (like anyons in $2d$ and flux-loops in $3d$) transform under translation symmetries. These anomalies are closely related to Lieb-Schultz-Mattis (LSM) type of theorems that constrain the possible low energy theories of a given lattice system\cite{LSM,Oshikawa,Hasting,Cheng_2016,ThorngrenElse,MetlitskiThorngren,ChoHsiehRyu}. We briefly describe the LSM-type of anomaly for topological orders in two and three dimensions in Appendix~\ref{LSManomaly}. These results reduce to the previously obtained boundary Luttinger relations in the absence of polarization, as described in Ref. \cite{KaneLuttinger} and also apply to $2$d systems with a nonzero Hall conductivity provided appropriate gauge choice in appendix \ref{square_numeric}, discussed previously in Ref.\cite{chern_band} within band theory.

%%%%%%%%%\subsection{Luttinger theorem in a dislocation}

The logic we used to study the boundary can also be used to study dislocations. A dislocation has space dimension $(d-2)$, and therefore can preserve at most $(d-2)$ translation symmetries. For simplicity we consider a dislocation with Burgers vector $\vec{B}=\hat{x_2}$ and unbroken translation symmetries $\mathbb{Z}^{x_3}\times...\mathbb{Z}^{x_d}$. The polarization term Eq.~\eqref{gaugeresponse} reduces to the following on the dislocation (with space-time dimension $d-1$):
\be
S_{dislocation}=P_1\int A\wedge x_3\wedge...x_d.
\ee
This has the same form as the boundary term Eq.~\eqref{polarizedcharge}, only in one dimension lower. This term leads to the same Luttinger theorem violation as Eq.~\eqref{LuttingerViolation} on the $(d-2)$ dimensional dislocation, where $V_F$ is again interpreted as the anomaly indicator of the low energy effective theory. A special case is $d=2$ which has been discussed in Ref.~\cite{DislocationCharge}, where the $V_F$ term is not needed and the polarization directly determines the fractional electric charge nucleated at the dislocation point.

\section{Summary}
\label{secconclusion}
 In this work we proposed an nonperturbative definition of the physically measurable polarization density in a crystalline insulator through translation properties of test magnetic monopoles. Our formalism is applicable  in any space dimension to  systems of interacting electrons but equally to  interacting bosons or spins that enjoy a $U(1)$ symmetry,  as long as there is a unique gapped ground state. The central result is a response involving background $U(1)$ fluxes and translation gauge fields, captured by a topological term. This response is topological despite the fact that the coefficient which is identified with the polarization is not quantized. 
%As opposed to the quantized Hall conductance, 
Indeed, to probe this response  
%nonquantized coefficients in Eq.~\eqref{gaugeresponse} identified as polarization, 
one needs to implement a global (non-perturbative) change, e.g. a $U(1)$ flux (monopole/instanton), a lattice shear, a dislocation or a physical boundary. This surprising connection seems natural in light of the necessity of charge quantization in order to properly define polarization\cite{vanderbilt_1993}, which is the consequence of a compact $U(1)$  symmetry group, which, by the  Dirac quantization argument, 
is related to the existence of magnetic monopole operators. The subtleties in previous literature associated with defining polarization ($\vec P$ vs $\tilde {\vec P}$) are neatly accounted for by gauge field configurations that have different distributions within a unit cell, corresponding to different approaches towards the continuum limit. Besides given a recipe to obtain the polarization in numerical calculations, the connection bears conceptual significance to boundary physics. We see that the classical relation between polarization and boundary charge density receives a quantum correction in the form of anomaly associated with the boundary low energy theory. For a boundary Fermi liquid the anomaly is associated with the familiar Luttinger theorem, and for a more general boundary phase, it is associated with a  Lieb-Schultz-Mattis like theorem, but for a general filling. 

\emph{Note added:} We note recent works\cite{Manjunath_2020} that studied the response of $(2+1)$D abelian topological phases with crystalline symmetries utilizing crystalline gauge fields for translations and rotations.

\section*{Acknowledgements}

  We gratefully acknowledge helpful discussions with G. Baskaran, Jing-Yuan Chen, Meng Cheng, Gil Young Cho, Dominic Else, Davide Gaiotto, Charlie Kane, T. Senthil, Yi-Zhuang You and Liujun Zou.  A.V. was supported by a Simons Investigator award and by a grant  from the Simons Foundation (651440, AV).  Research at Perimeter Institute (YCH and CW) is supported by the Government of Canada through the Department of Innovation, Science and Economic Development Canada and by the Province of Ontario through the Ministry of Research, Innovation and Science. 

\appendix
\section{Polarization with nontrivial unit cell structures}
\label{PPtilde}

Here we discuss different definitions of polarization  to emphasize the lattice point of view. Most of the physics below are discussed in Refs.~\cite{WatanabeOshikawa,Bardarson_2016}. 

To illustrate the essential point, it suffices to consider a simple $1d$ lattice with two sub-lattices and one electron orbital on each -- generalizations to more complicated unit cells (or even higher dimensions) will be straightforward. We label the unit cells by $i\in\mathbb{Z}$ and sub-lattices by $\{a,b\}$. We consider a simple insulator with unit charge occupation per unit cell, namely $\langle n_{i,a}+n_{i,b}\rangle=1$, with an extra unit negative (ion) charge $q_{ion}=-1$ sitting on site $a$ to make the entire system charge-neutral.

The electric polarization is defined via the response to a smooth gauge field $(\varphi,A)$ defined on the lattice -- but what does ``smooth" mean? Clearly we want the gauge field to be slowly varying when moving from one unit cell to another, which means the lattice momentum of the gauge field is close to zero. For example $A_{i,a;i,b}\approx A_{i+1,a;i+1,b}$ and $\varphi_{i,a}\approx\varphi_{i+1,a}$. However this does not uniquely specify how the gauge field should be distributed within a unit cell, namely how  $A_{i,a;i,b}$ should be compared with $A_{i,b;i+1,a}$, or how $\varphi_{i,a}$ should be compared with $\varphi_{i,b}$.

Apparently we have a choice to make here. One simple choice is to demand $A_{i,a;i,b}=0$ and $\varphi_{i,a}=\varphi_{i,b}$. {The continuum limit of the gauge fields (now a smooth function of the continuum coordinate $x$ with no explicit dependence on the sublattice index) will be $A(x=i)=A_{i,b;i+1,a}$ and $\varphi(x=i)=\varphi_{i,a}$.} This is equivalent to viewing the entire unit cell as a single point in space. The polarization defined via the response to such field configurations effectively measures only the inter-unit cell entanglement, and is called $\tilde{\vec{P}}$ in Ref.~\cite{WatanabeOshikawa}.
%\ashvin{Can we be a bit more explicit here? what is the choice of vector potential presumably $A_{i,b;i+1,a} = something$ so it clearly is the inter unit cell entanglement? Also can we specify the functional derivative we need to do? }

We can make a more general choice as follows: we demand
\bea
\label{Platticegauge1}
(1-\alpha)A_{i,a;i,b}&=&\alpha A_{i,b;i+1,a}, \nn (1-\alpha)(\varphi_{i,b}-\varphi_{i,a})&=&\alpha(\varphi_{i+1,a}-\varphi_{i,b}),
\eea for some constant $\alpha$. {The continuum limit is then 
\bea
\label{Platticegauge2}
A(x=i)&\equiv&\alpha A_{i,a;i,b}+(1-\alpha)A_{i,b;i+1,a}, \nn 
\varphi(x=i)&\equiv&\varphi_{i,a}.
\eea } 
The physical meaning is also clear: we interpret the two sublattices $a,b$ as separated by a distance $\alpha$ in real space (with lattice unit normalized to unity). If the lattice system originates from a continuum with the two sites separated physically by a distance $\alpha$, then this choice corresponds to a physically uniform electric field, 
%\ashvin{(HERE do we impose both scalar and vector potentials? it would again be helpful to know what the function derivative involved is in terms of the probe fields eg. can we set $\varphi_{i,a} = E i $ and $\varphi_{i,b}= E (i=\alpha)$) and vary with respect to E?)} 
and therefore produces the polarization $\vec{P}$ for a uniform electric field. To obtain the polarization difference $P-\tilde{P}$, consider the lattice-scale action difference $S[A^{\mu}_P]-S[A^{\mu}_{\tilde{P}}]$, where $A^{\mu}_P,A^{\mu}_{\tilde{P}}$ are the corresponding lattice gauge field distributions with the same continuum limit $A^{\mu}(x,t)$. Using the definition of $A^{\mu}_P$ in Eq.~\eqref{Platticegauge1},\eqref{Platticegauge2} and the fact that $j_{i,a;i,b}-j_{i,b;i+1,a}=\partial_t n_{i,b}$ where $j$ is the lattice current operator that couples to $A$ field, one can see that 
\bea
&&S[A^{\mu}_P]-S[A^{\mu}_{\tilde{P}}]\nn
&&=\sum_i\int dt \,\alpha n_{i,b}[\partial_t A(x=i)-(\varphi(x=i+1)-\varphi(x=i))], \nn
\eea
which simply means that $\vec{P}-\tilde{\vec{P}}=\alpha\langle n_{i,b}\rangle$, and can be interpreted as a classical contribution, namely a charge $q_b$ sitting at site $b$ contributes a dipole moment $\alpha q_b$.
%} \ashvin{again explicit functional derivative would help here, I could not see this}.  %We also emphasize that even in this case we are still perfectly allowed to ask about the response of the system to an electric field non-uniform within a unit cell, so other choices of gauge field distribution within a unit cell can still be legitimate depending on context. 

Different choices of the intra-cell gauge field distribution also lead to different definitions of charge density (per unit cell) in the continuum limit. In the continuum limit we define charge density as $\rho(x)=\delta S/\delta\varphi(x)$ {where $\varphi(x)$ is in the continuum limit as discussed above}. This is easy for the gauge field probing $\tilde{\vec{P}}$, where $\varphi$ is unique within a unit cell, and we simply get $\rho_i=\langle n_{i,a}+n_{i,b}\rangle+q_{ion}$. However with nontrivial $\alpha$ (as for the standard $\vec{P}$), there is a correction from the non-uniformity of $\varphi$ within the unit cell. A simple calculation gives $\delta\rho_i=-\alpha\nabla\langle n_{i,b}\rangle$, which is nonzero only at the boundary. The relation $\rho=-P$ (mod $1$) at the  boundary (an {insulating} boundary in $d>1$)holds for both $P$ and $\tilde{P}$ as long as the corresponding definitions for charge density are used. {A similar difference between current operators from $j(x)=\delta S/\delta A(x)$ in different continuum limits was also discussed in the literature.\cite{WatanabeOshikawa}}
%\ashvin{This is for Insulating boundaries - we need to specify right?} {\color{purple}(Yes although in this $1d$ example it's automatic, but I've added the word "insulating" above to avoid confusion.)}

The above discussion implies that when one considers small momentum (long wavelength) components of $dA$, e.g. a uniform electric field induced by monopole translation, Eq.~\eqref{gaugeresponse} gives a response controlled by the conventional polarization $\vec P$; while $dA$ around higher momenta $n_i\vec G_i (n_i\in \mathbb Z)$ probes sites inside a unit cell with different weights. In one extreme case, where one concentrates inductive electric field on only inter-cell bond, the response term gives $\vec{\tilde P}$.

\section{Details of 2d calculation}
\label{2DNumerics}

\subsection{Recipes for calculation}
\label{translation_berry}

Consider a square lattice with $L_x\times L_y$ unit cells, %Since we are neglecting lattice symmetries other than translations, any $2d$ lattice can be represented this way. 
and assume a unique gapped ground state. We put a total magnetic flux of $2\pi$ uniformly on the entire surface. To be concrete, let us take the following gauge (analogue of Landau gauge on a discrete torus):
\bea
\label{LandauGauge}
A_{i,i+\hat{x}}&=&-\frac{2\pi y}{L_y}\delta_{x,L_x-1}, \nn
A_{i,i+\hat{y}}&=&\frac{2\pi x}{L_x L_y}, \nn
i=(x,y), & & x\in\{0,...L_x-1\}, y\in\{0,...L_y-1\}. 
\eea
In this gauge a unit translation in $\hat{x}$ (denoted $T_x$) should be followed by a gauge transform that acts nontrivially only on the $x=0$ strip:
\bea
G_x={\rm exp}\left(-i\sum_i\frac{2\pi y}{ L_y}\delta_{x,0}\hat{q}_i \right),
\eea
where $\hat{q}_i$ is the charge density operator on site $i$. The $\hat{y}$-translation (denoted $T_y$), in contrast, does not need an additional gauge transform.

Strictly speaking, however, on a finite torus neither $G_xT_x$ nor $T_y$ is a true symmetry since the Wilson loop along the nontrivial $\hat{y}$ and $\hat{x}$ cycles cannot be translationally invariant under $T_x$ and $T_y$, respectively. As we can see explicitly from Eq.~\eqref{LandauGauge}, the Wilson loops changes by $\int dy \delta A_y=-2\pi/L_x$ on every $\hat{y}$-cycle and $\int dx \delta A_x=2\pi/L_y$ after $T_x$ and $T_y$, respectively. This non-invariance of Wilson loops cannot be cured by a gauge transform. To overcome this issue, we consider modified translations $\tilde{T}_x=F_y G_xT_x$ and $\tilde{T}_y=F_xT_y$, where $F_y, F_x$ are adiabatic evolutions that modify the $A$ fields by $\delta A$ at the end of the evolutions, where
\bea
F_y: && \hspace{10pt} \delta A_x=0, \hspace{10pt} \delta A_y=\frac{2\pi }{L_xL_y},\nn
F_x: && \hspace{10pt} \delta A_x=-\frac{2\pi}{ L_y}\delta_{x,L_x-1}, \hspace{10pt} \delta A_y=0. 
\eea
The composite operations $\tilde{T}_x, \tilde{T}_y$ preserve the Hamiltonian, and therefore produce well-defined Berry phases $\phi_x,\phi_y$ which we identify with $\vec{k}_e=-\vec{k}_{\mathcal{M}}=2\pi(P_y,-P_x)$.

The connection between the monopole momentum and polarization can also be understood from the structure of $\tilde{T}_x, \tilde{T}_y$. Consider the operations $\tilde{T}_x^{L_x}$ and $\tilde{T}_y^{L_y}$. Using $T_x^{L_x}=T_y^{L_y}=1$, one can see that the two operations become the familiar adiabatic $2\pi$ flux threading in the $\hat{y}$ and $-\hat{x}$ directions, respectively. The corresponding Berry phases are $(\Phi_x, \Phi_y)=2\pi (L_xP_y,-L_yP_x)$, in agreement with our previous result. Notice that since $\tilde{T}_i^{L_i}\neq1$, the translation Berry phase defined above is not quantized on a finite system -- this is consistent with the fact that polarization can take continuous value in a finite system.

In practice, since $F_x, F_y$ only threads a small flux of order $O(1/L)$, one would expect their actual effect to be small, especially at large $L$. One can then consider the simpler amplitudes $\la\Omega|G_xT_x|\Omega\ra$ and $\la\Omega|T_y|\Omega\ra$ ($|\Omega\ra$ being the ground state in the flux background). These will have magnitudes smaller than one on a finite torus, but as long as it is non-vanishing (in fact we expect it to approach unity in the thermodynamic limit), one can extract the phase of the amplitude, and this phase should give the monopole momentum, which in turn gives the polarization density. More explicitly
\be
\label{Restastar}
\la\Omega|G_xT_x|\Omega\ra=\rho_xe^{2\pi i P_y}, \hspace{10pt} \la\Omega|T_y|\Omega\ra=\rho_ye^{-2\pi iP_x},
\ee
where $\rho_{x,y}$ are magnitudes that are non-vanishing in the thermodynamic limit (in practice they $\to 1$, see Sec.~\ref{Example}).  Eq.~\eqref{Restastar} is in the same spirit with Resta's formula\cite{Resta} for polarization in one dimension, which is the phase of the (smaller than one) amplitude $\la\Omega|\exp(ix\hat{q}_x/L)|\Omega\ra$. In higher dimensions Resta's amplitude vanishes in the thermodynamic limit and cannot be used to extract polarization\cite{WatanabeOshikawa}. Our prescription using the amplitudes $\la\Omega|G_xT_x|\Omega\ra$ and $\la\Omega|T_y|\Omega\ra$ can be viewed as a proper generalization of Resta's formula to two dimensions. In fact this prescription has been carried out in previous studies of monopoles in two-dimensional $U(1)$ spin liquids \cite{monopole2}.

Strictly speaking our recipe gives the polarization of the ground state in the $2\pi$-flux background $|\Omega\ra$, which is slightly different from the original ground state without the flux $|\Omega\ra_0$. The two should agree in the thermodynamic limit. To see this let us consider insulators with zero Hall conductance. If there is no symmetry other than charge conservation and translations, the leading order term in the response theory that can cause a magnetic flux to change the polarization is $\Delta\mathcal{L}\sim \alpha_iBE_i$ for some constants $\alpha_i$ ($i=x,y$). This means that a total $2\pi$-flux will change polarization by $O(B)\sim O(1/L^2)$. With time-reversal symmetry the leading order term becomes $\sim B^2E$, and the change of polarization in the $2\pi$-flux background becomes $O(B^2)\sim O(1/L^4)$. This error will likely be dominated by other finite-size effects such as omitting the flux-threading $F_{x,y}$ in the calculation. This argument is reliable for insulators without Hall conductance since we expect all terms in the response theory to be local and manifestly gauge-invariant. %Even for insulators with Hall conductance (where the polarization is not gauge invariant), the Chern-Simons response term $AdA\sim \vec{A}\times \vec{E}$ is more singular, and by simple power counting we conclude that the change of polarization density in the $2\pi$-flux background is $\sim O(1/L)$} %In any case the flux background induces a change in polarization that is vanishing in the thermodynamic limit. %({\color{red}CW: I'm tempting to say that even for Dirac fermions the change in polarization due to the flux background is $O(1/L^4)$ if other finite-size effects are suppressed, since the non-local response terms should be conformally symmetric. But let me think a bit more...})

If the unit cell contains $2$ neighboring sites in $x$ direction, i.e. $L_x$ even and a unit cell contains $(2n,m),(2n+1,m) (n,m\in \mathbb Z)$, the above recipe only distributes non-vanishing $A$ field on bonds between unit cells of choice, and $A$ vanishes within each unit cell, which corresponds to calculating $\vec{\tilde P}$ in Ref.~\cite{WatanabeOshikawa}. The actual unit cell structure and geometry do not contribute to monopole translation properties, or polarization, obtained in such ways. In general the polarization and the monopole momentum depend on choice of unit cells.

To obtain the polarization $\vec P$, with both intra-(classical) and inter-cell effects, we give a recipe to account for the unit cell geometry, applicable to generic systems. To this end we first give a continuum function for gauge field $\vec A$ on the torus which can then be used to determine the discrete gauge fields. Take the distance between neighboring unit cells to be $1$ and the Bravais lattice to be square, the continuum gauge field reads
\begin{eqnarray}
\label{continuum}
A_x(x,y)=\begin{cases} 0 & 0\leq x<L_x-1 \\ -\frac{2\pi y}{ L_y} & L_x-1\leq x<L_x\end{cases}\nonumber \\
A_y(x,y)= \begin{cases} \frac{2\pi x}{L_xL_y} & 0\leq x\leq L_x-1  \\  \frac{2\pi( L_x-1)}{L_xL_y} (L_x-x) & L_x-1<x<L_x \end{cases}.
\end{eqnarray}
Note the function is not single-valued, but is well defined and hence poses no problems for obtaining the gauge fields on the discrete lattice. When put on the lattice, the gauge connection on one bond $l$ is given by $\int_l d\vec x \cdot \vec A(\vec x)$, i.e., the line integral of continuum $\vec A$ along the bond. 

Once put on a lattice, the flux close to the ``slit" at $y=L_y, L_x-1\leq x\leq L_x$ should have an $O(1)$ deviation from $2\pi/(L_xL_y)$ due to the discontinuity in eq \eqref{continuum} (the total flux threading the unit cell at $(L_x-1,L_y-1)$ is hence $2\pi/(L_xL_y)-2\pi$). One could compensate for this deviation by altering the gauge connection on bonds inside the slit, such that the deviation is concentrated to a set of elementary plaquettes (i.e., not containing any smaller plaquettes) that contain the point $(L_x,L_y)$, whose flux equals $2\pi/(L_xL_y)A-2\pi$ ($A$ is the area of the elementary plaquette). This fixes the translation symmetry breaking of flux derived from the continuum recipe. Upon translation $T_y$, one carefully performs a gauge transform on sites in the slit $G_y$ to restore the gauge connection as much as possible, the amplitude of $\langle \Omega |G_yT_y|\Omega\rangle$ is comparable to unity; the phase converges in thermodynamic limit to the conventional polarization $\vec P$.

The two recipes have the same flux configuration on torus and hence are connected by a gauge transform. However, this gauge transform generally does {\em not} commute with $G_yT_y $ and will change the momentum obtained, consistent with getting $\tilde {\vec P}$ versus $\vec P$ for the two recipes. For example, in the above square lattice model, we assume a unit cell at $(n,m)$ contains two sites at $(n,m),(n+1/2,m)$, respectively (in notation of eq ~\eqref{LandauGauge} the coordinates read $(2n,m),(2n+1,m)$). Then the two recipes built upon eqs~\eqref{LandauGauge} and \eqref{continuum} differ by a gauge transform on sites with $x=L_x-1/2$ by the operator $e^{-i\sum_y \frac{\pi y}{ L_y}\hat\rho(L_x-1/2,y)}$. From the commutation relation between this gauge transform and the $G_yT_y$ operation, one can see that the change of momentum from the gauge transform as $L\to\infty$ is precisely $\delta k_y=-\pi\langle\hat{\rho}(L_x-1/2,y)\rangle$, which leads to a change in polarization $P_x-\tilde{P}_x=\langle\hat{\rho}(L_x-1/2,y)\rangle/2$, in agreement with the intuition that the difference between $\vec{P}$ and $\tilde{\vec{P}}$ can be seen as a classical dipole moment within the unit cell. 

\subsection{Review of band theory calculation}
\label{Dirac}

For a $d$-dimensional lattice system with translation symmetries and periodic boundary conditions in all directions, the polarization corresponds intuitively to the dipole moment in each unit cell.  For free fermions the polarization  contributed by an occupied band is given by the integrated Berry connection (the Wilson loop) in the Brillouin zone\cite{vanderbilt_1993}: 
\be
\label{PBerry}
\vec{P}=\int_{BZ} \frac{d^d \vec{k}}{(2\pi)^d}\la u_{\vec{k}}|i\partial_{\vec{k}}|u_{\vec{k}}\ra \hspace{10pt} ({\rm{mod}}\hspace{2pt} 1),
\ee
where $|u_{\vec{k}}\ra$ is the periodic part of the Bloch state at momentum $\vec{k}$ and the integration is taken over the entire Brillouin zone. (Here $u_{\vec k}(r)=e^{-i\vec k\cdot \vec r} \psi(r)$.)
We also discuss $\tilde {\vec P}$ \cite{WatanabeOshikawa} if we instead use $\tilde u_{\vec k}(\vec r)=e^{-i\vec k\cdot \vec R} \psi(\vec r)$ where $\vec r=\vec R+\vec r_i$ and $\vec R$ is the Bravais lattice vector associated with $\vec r$. ($\tilde u_{\vec k}(\vec r)=\tilde u_{\vec k+\vec G_i}(\vec r)$.)

In the presence of gapless Dirac cones, the band theory polarization Eq.~\eqref{PBerry} is not uniquely defined. This ambiguity can also be understood from monopole momentum: in a $2\pi$-flux background there are fermion zero modes associated with the Dirac fermions, and filling different zero modes gives different ground states, with different total momenta. We now discuss this within the usual band theory formulation. For concreteness consider a system of spin-$1/2$ fermions forming two Dirac valleys, say at momenta $\vec{K}, \vec{K}'$. The Wilson loop for each spin $\alpha$ in the $\hat{k_1}$ direction
\be
\mathcal{P}_{\alpha}(k_2)=\int \frac{dk_1}{2\pi}\la u_{\alpha,\vec{k}}|i\partial_{k_1}|u_{\alpha,\vec{k}}\ra 
\ee
has a discontinuity of $\pm\pi$ when $k_2$ passes through $K_2$ and $K'_2$. The polarization
\be
P_1=\sum_{\alpha=\uparrow,\downarrow}\int\frac{dk_2}{2\pi}\mathcal{P}_{\alpha}(k_2)
\ee
requires a choice of the jump in $\mathcal{P}_{\alpha}$ ($\pi$ or $-\pi$) at each Dirac point. In order for the polarization to be gauge-invariant, $\sum_{\alpha}\mathcal{P}_{\alpha}(k_2)$ should be single-valued in the entire Brillouin zone (i.e. no net Chern number). This leads to six different choices of the jumps in $\mathcal{P}_{\alpha}$ at the Dirac points. Now from the monopole momentum point of view, in a $2\pi$-flux background there are four zero modes (one from each Dirac cone), and gauge-invariance requires the ground state to fill half of the zero modes, which leads to $C^4_2=6$ different choices -- in exact agreement with the band theory consideration.

\subsection{An example}
\label{Example}

Our numerical prescription for calculating polarization density through amplitudes like $\la\Omega|T_y|\Omega\ra$ ($|\Omega\ra$ being the many-body ground state in the presence of a uniform $2\pi$ flux background) is well-defined for generic many-body systems. In the special case of free fermions we expect our prescription to agree with the band theory results from Eq.~\eqref{PBerry}. We demonstrate this through an example of a Dirac semimetal (with a specific choice of zero-mode fillings). We consider a square lattice, labeled by two orthogonal unit lattice vectors $e_{1,2}$, with $2$ orbitals and $2$ spin species on each site. The Hamiltonian for our spin-$1/2$ fermions reads
\begin{equation}
    \mathcal H=\sum_{\langle ij\rangle,\alpha,\beta} t_{i\alpha,j\beta}e^{i a_{ij}} f^\dagger_{j\beta,s}f_{i\alpha,s}+\sum_{i,\alpha,\beta} t_{i\alpha,i\beta} f^\dagger_{i\beta,s}f_{i\alpha,s}
\end{equation}
where $s=\uparrow,\downarrow$ labels spin indices, $\alpha,\beta=1,2$ label orbitals,$\langle ij\rangle$ denotes neighboring or sites linked by a diagonal bond and hopping amplitudes read
\begin{eqnarray}
\label{square_hopping}
t_{[l_1,l_2],[l_1+1,l_2]}&=&1\nonumber\\
t_{[l_1,l_2],[l_1,l_2+1]}&=&(-1)^{l_1}\nonumber\\
t_{[l_1-1,l_2],[l_1,l_2-1]}&=&(-1)^{l_1}t \quad (t\in [0,1])\nonumber\\
t_{[l_1,l_2][l_3,l_4]}&=&t_{[l_3,l_4][l_1,l_2]}
\end{eqnarray}
where we have relabeled subscripts $i\alpha$ by $[l_1,l_2]$ through $l_1=2*i_1+\alpha,l_2=i_2$ (site $i$ with coordinates $(i_1,i_2)$ in $e_{1,2}$ basis) and $t$ is the tuning parameter. Hopping amplitudes on other diagonal or neighboring bonds not covered in eq \eqref{square_hopping} vanish. The two limits $t=0,1$ correspond to a square with $C_4$ rotation and an effective triangular lattice with $C_6$ rotation, respectively. Diagonalizing this Hamiltonian in momentum space gives gapless dispersion at half-filling. To avoid the ambiguity for Wilson loop operator when crossing Dirac fermions as discussed in sec \ref{Dirac}, we stipulate the two bands for spin up/down has Chern number $\pm 1$, respectively, i.e. effectively open an infinitesimal quantum spin hall mass. For the monopole momentum calculation, the gauge connection on links $a_{ij}$ analogous to Landau gauge in eq \eqref{LandauGauge} gives a total flux of $2\pi$ and the quantum spin hall mass indicates that one fills only $2$ zero modes of one spin species, giving a monopole carrying spin $1$. Fig \ref{fig:numerict} in the main text shows a comparison of polarization $P_1$, calculated numerically using eq \eqref{PBerry} along the direction of reciprocal vector for $e_1$ and the monopole momentum $k_2$ along the orthogonal $e_2$ direction, calculated as in sec \ref{translation_berry} as one tunes $t$ from $0$ to $1$. The polarization obtained from Eq.~\eqref{PBerry} is discretized as summation of Berry phase $\ln \langle u_{\mathbf k} | u_{\mathbf {k+\epsilon}}\rangle$ for $30000$ points in Brillouin zone; the momentum is calculated on a lattice of linear size $L=50$. For the special cases when $t=0,1$ the momentum $k_2=\pi,2\pi/3$ agrees with results in Refs.~\cite{monopole1,monopole2}.

\section{Polarization and other topological quantities}
\label{HallTheta}

In $2d$ the polarization density is not invariant under large gauge transforms if the system has a nonzero Hall conductance $\sigma_{xy}\neq0$. This is known in band theory, where Eq.~\eqref{PBerry} is not invariant under large gauge transforms in real space -- in fact this is one way to define integer quantum Hall effect within band theory. Beyond band theory, it is also easy to understand why this is so from the monopole momentum: a $2\pi$-flux induces an extra charge $\delta Q=\sigma_{xy}$ in the ground state, which makes the total momentum non-invariant under large gauge transforms. The total polarizations $P_xL_y$ and $P_yL_x$ are still well-defined (gauge invariant) mod $1$. Similarly, if the system forms a quantum spin Hall insulator, with a nonzero $S_z$ spin trapped in a magnetic flux unit, then the polarization is not invariant under a large $S_z$-gauge transform. In all such cases the polarization remains meaningful (unambiguisly defined) for a given gauge if the gauge field remains non-dynamical.

Contrary to the Hall conductance, a nonzero magnetoelectric angle $\Theta$ in $3d$:
\be
\label{Theta}
\frac{\Theta}{4\pi^2}\vec{E}\cdot\vec{B}
\ee
does not obstruct the gauge-invariance of polarization density. Within band theory the $\Theta$-angle can be interpreted as the magnetoelectric polarizability\cite{QiHughesZhang, EssinMooreVanderbilt}, i.e. a magnetic field induces an extra polarization density
\be
\label{Magnetoelectric}
\Delta \vec{P}=\frac{\Theta}{4\pi^2}\vec{B}.
\ee
The monopole point of view provides a simple understanding of the above relation beyond band theory: when $\Theta\neq0$, the monopole traps a fractional charge $q=\Theta/2\pi$ and becomes a dyon\cite{WittenDyon}. When a magnetic field is turned on, say in $\hat{z}$, the monopole also sees the field due to the fractional $q$. This contributes to the non-commutativity of $T_x$ and $T_y$, with the additional phase factor given by $qB=(\Theta/2\pi)B$. Using Eq.~\eqref{3dMonopole} as the definition of polarization we immediately obtain Eq.~\eqref{Magnetoelectric}.

\section{Anomaly from a Fermi surface}
\label{FSAnomaly}
In this appendix, we derive the anomaly term Eq.~\eqref{Anomaly} which proves Luttinger theorem in any spatial dimension $d$. The logic is to partition the Fermi surface into infinitesimally small patches in whose proximity reside ``chiral" fermions, that effectively live in $(1+1)d$. The chiral anomaly from each of these fermions adds up to give Eq.~\eqref{Anomaly}.

Let us first write down the anomaly for a $1d$ chiral fermion with the free Hamiltonian $\psi^\dagger i(\pm\partial_x) \psi$, where we set velocity to unity and $\pm$ represents right(left)-movers under consideration. Next we couple to theory to both a $U(1)$ electromagnetic field $A$ and an x-translation gauge field (elasticity tetrad) $x$. The momentum of the chiral fermion $k_F$ becomes the coupling constant between the translation gauge field $x$ and the fermion (in analogy to electric charge $e$ as the coupling constant between EM field $A$ and a fermion, here the charge of translation - momentum - mediates the coupling ). Hence the covariant derivative $ i\partial_{x,t}\rightarrow  i\partial_{x,t}+A_{x,t}+ k_F x_{x,t}$ where subscript denotes the space-time component of the $1-$ form gauge field, omitted hereafter.  To obtain the mixed anomaly between $A,x$, one goes to one higher dimension $(2+1)d$ bulk of the chiral fermion - a quantum hall insulator with Chern number $C=\pm 1$, with the low-energy topological quantum field theory action 
\begin{equation}
    \label{cs}
S=\frac{\pm 1}{4\pi}\int (A+ k_Fx)\wedge d(A+ k_Fx).
\end{equation}

Note that since we introduce elasticity tetrads in addition to $A$, the familiar Chern-Simons term $A\wedge dA$ is modified as such. The mixed term in Eq.~\eqref{cs} reads $\pm\int\frac{k_F}{2\pi}x\wedge dA$ from which descends a boundary term $\mp \int\frac{k_F}{2\pi} A\wedge x$. 

Now that we have the desired $(1+1)d$ anomaly, consider in $d$ space dimension system, compactify $(d-1)$ dimensions and derive similar anomaly for the effective $(1+1)d$ system along the remaining $i$th primitive lattice vector direction. We inspect a small patch on the Fermi surface with momentum range $(k_1\pm \delta k_1/2,\cdots k_i,\cdots k_d\pm \delta k_d/2)$ ($\delta k_j\geq 0$,the variation of $k_i$ is neglected to zeroth order of the anomaly). On such a patch with an ``area" $\Delta s_i = \prod_{j\neq i} \delta k_j$, there are $\left(\frac{\Delta s_i}{(2\pi)^{d-1}} \prod_{j\neq i} L_j\right)$   chiral fermions along the $i$th direction, each associated with an anomaly $\frac{ \mp k_i}{2\pi} \int A\wedge x_i$ ($\mp$ in numerator results from right(left)-movers given by the orientation of the small patch projected onto $i$th reciprocal vector direction). Adding all patches up, the self Chern-Simons terms vanish and the remaining anomaly reads
\begin{eqnarray}
\label{fsanomaly}
S_{FSanomaly}=-\frac{\prod _{j\neq i} L_j}{(2\pi)^d} \sum_{FS} \Delta s_i \eta_i k_i\int A\wedge x_i\nonumber\\\rightarrow -\frac{V_F}{(2\pi)^d} \int A\wedge \prod x_i.
\end{eqnarray}
where $\sum_{FS}$ counts all patches on the Fermi surface, $\eta_i=\pm 1$ denotes the orientation of each patch along/against $i$th reciprocal lattice vector and we use the identity on luttinger volume $\sum_{FS} \Delta s_i \eta_i k_i= V_F$. The second line arises after we introduce translation gauge fields along the other $(d-1)$ directions and the numerator in the first line $\prod_{j\neq i} L_j\rightarrow \int \wedge \prod_{j\neq i} x_j$. The final result puts all $x_i$'s on equal footing and hence it correctly captures the anomaly of Fermi surface under large gauge transforms along any spatial directions. Adding the anomaly term to the Fermi surface theory will make the full theory anomaly-free, as promised in Eq.~\eqref{SFull}.

\section{LSM anomaly indicators for topological orders}
\label{LSManomaly}

 First we review an important notion for a topological order with a global $U(1)$ symmetry in general $d$ dimensions known as the fluxon. Consider an instanton of the $A$ field, which is an operator supported on a $(d-2)$ dimensional sub-manifolds in space, with $\int dA=2\pi$ on the two complementary spatial dimensions. For $d=2$ it is a point flux insertion and for $d=3$ it is a unit flux loop. Dirac quantization requires this object to be un-observable from far away. However in a topologically ordered state, there can be nontrivial quasiparticles that carry fractional electric charge, and moving these fractional charges around the instanton will naively produce an observable Aharanov-Bohm phase. The resolution is that the bare instanton is attached with another nontrivial excitation, called the fluxon, from the topological order. The property of the fluxon is such that the combined object becomes unobservable from far away. For example, a fractionally charged quasiparticle will have a nontrivial braiding phase with the fluxon so that it braids trivially with the combination of fluxon and bare instanton. In $2d$ the fluxon is an anyon excitation and in $3d$ it is a loop excitation.

In general, anomalies involving a $U(1)$ global symmetry in topological quantum field theories are encoded in the properties of fluxons. Essentially if the fluxon carries a fractional quantum number under other symmetries, in our case lattice translations, then the instanton  will also carry the fractional symmetry quantum numbers since it is bound with a fluxon. Since the instanton is supposed to be unobservable, this becomes an anomaly. The fluxon has space-time dimension $d-1$, and crystal symmetry fractionalization can be described using a partition function in $d$ space-time dimension:
\be
\label{FluxonPSG}
\mathcal{L}_{Fluxon}=-n_A\int x_1\wedge x_2...\wedge x_d,
\ee
for which the fluxon lives on the boundary of the $d$-dimensional (space-time) manifold, and $n_A\in[0,1)$ is the LSM anomaly indicator.

At $d=2$ the fluxon is an (abelian) anyon particle, and Eq.~\eqref{FluxonPSG} means that the fluxon transforms projectively under translation symmetries:
\be
\label{fluxonflux}
T_2^{-1}T_1^{-1}T_2T_1=e^{-i2\pi n_A}.
\ee
This relation has been discussed in Ref.~\cite{LuRanOshikawa}. We note that this result is equally applicable for magnetic translation symmetries, where a nontrivial $U(1)$ flux $\phi$ is enclosed in each unit cell. As a simple example, consider a short-range entangled integer quantum Hall state. The ``fluxon'' in this case must be an integer multiple of the local electron since there is no fractional excitation. Specifically, to make the $2\pi$-flux unobservable, the fluxon must carry electric charge $-2\pi\sigma_{xy}$, namely it is the bound state of $-2\pi\sigma_{xy}$ electrons. The effective magnetic flux seen by this fluxon is therefore $-2\pi\phi\sigma_{xy}$. To satisfy Eq.~\eqref{fluxonflux} we must therefore have
\be
\phi\sigma_{xy}=n_A\hspace{5pt}({\rm{mod}}\hspace{2pt}1).
\ee
This relation has also been discussed in Ref.~\cite{LuRanOshikawa}.

At $d=3$ the fluxon is a loop excitation with finite tension. Eq.~\eqref{FluxonPSG} has the following interpretation. First consider a straight fluxon tube, say pointing in $\hat{z}$ (assuming periodic boundary condition). If we translate the entire loop in the $(x,y)$ plane, $T_x$ and $T_y$ will commute only up to a phase: $T^{-1}_yT^{-1}_xT_yT_x=\exp(-2\pi i n_AL_z)$ where $L_z$ is the number of layers of the entire system in $\hat{z}$. Another way to describe this property, without relying on having a finite $L_z$, is to consider a closed fluxon loop that links with a dislocation (a line defect in $3D$), say with Burgers vector $\hat{z}$. Translation symmetries will act on the loop projectively. More generally, for a fluxon loop linked with a dislocation with Burgers vector $\vec{B}$, we have
\be
T^{-1}_jT^{-1}_iT_jT_i=\exp(-2\pi i\epsilon^{ijk}B_k n_A).
\ee
Another consequence, following similar reasoning, is a nonabelian three-loop braiding\cite{WangLevin} for a fluxon loop and two dislocations.

\section{ Square model numerics to verify boundary Luttinger theorem}
\label{square_numeric}

\begin{figure*}
 \captionsetup{justification=raggedright}
    \centering
   { \includegraphics[width=0.9\textwidth]{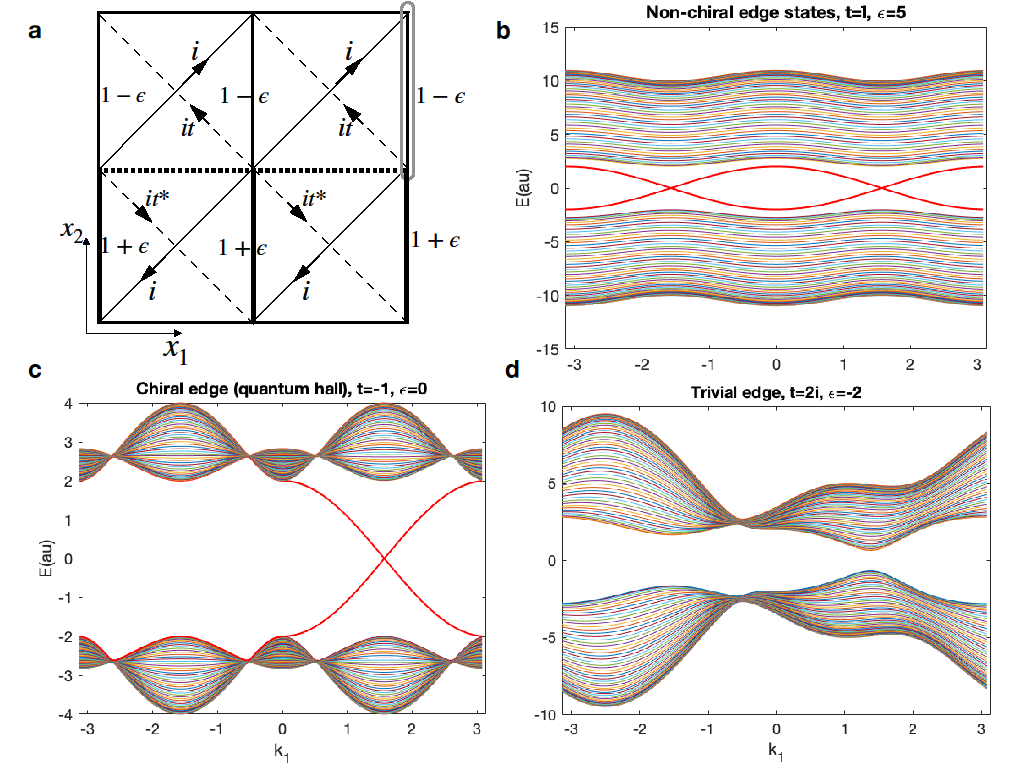}}
    \caption{Figure (a) illustrates the fermion hopping parameters used. The hopping strengths on solid/dotted horizontal bonds are $\pm 1$, on normal weight/bold vertical bonds are $1\mp \epsilon (\epsilon \in \mathbb R)$ and across solid/dashed diagonal bonds are $i,it^{(*)} (t\in \mathbb C)$ (direction denoted by arrows), respectively. The unit cell is doubled along vertical direction. (b)(c)(d) show typical energy spectrum on a cylinder geometry with boundaries at $x_2=0,L$ and periodic in $x_1$ direction. The in-gap red lines depict states localized at two edges. (b)(d) satisfy eq \eqref{LuttingerViolation} while (c) has non-vanishing Chern number and gauge-dependent polarization, boundary $k_F$, so boundary Luttinger theorem applies with an appropriate Berry connection integral rule.}
    \label{square_model}
\end{figure*}

Our square lattice model consists of spinless fermions with nearest neighbor and diagonal hopping, detailed configuration shown in Fig.~\ref{square_model}a. It's modified from $\pi$-flux square hopping with $\epsilon$ variation of vertical hopping, time-reversal breaking imaginary diagonal hopping and complex diagonal hopping $t$ further breaks remaining rotation (inversion), reflection symmetries to allow a generic polarization. The only symmetries that remain preserved is translation. 

When put on a cylinder geometry with $x_2=0,L$ boundaries and periodic along $x_1$ direction, we can calculate boundary charge densities for e.g. $x_2=0$ boundary as \cite{vanderbilt_1993}
\begin{eqnarray}
\label{bd_charge}
\rho_{bd}=\frac{1}{\Omega}\int_{-\infty}^{x_0} dx_2 \int_{x_2-a/2}^{x_2+a/2} dx'\int_{A_{bd}} dx_1\rho(x_1,x')
\end{eqnarray}
 where $\rho(x_1,x_2)$ is the charge density including ions (for a neutral system), $x_0$ locates deep in the bulk, $a$ is size of unit cell along $x_2$ direction, $A_{bd}$ denotes any segment covering exactly one unit cell on the boundary and $\Omega$ the unit cell area. This amounts to first averaging charge density $\bar \rho(x_2)=\frac{1}{\Omega}\int_{x_2-a/2}^{x_2+a/2} dx'\int_{A_{bd}} dx_1\rho(x_1,x')$ within a window $[x_2-a/2,x_2+a/2]$ to smoothen any irrelevant periodic oscillations in the bulk ($\bar \rho(x)=0$ for $x$ inside the bulk) while retain the extra charge accumulation \cite{Kudin_2007}, then integrating the averaged density. From the field-theoretic point of view, this window function $\bar\rho(x)$ use comes naturally from the application of the long-wavelength limit in eq \eqref{gaugeresponse} to discrete lattices. In continuum, one identifies each unit cell with a single point $\vec x$ and hence the vector potential $A_0(\vec x)$ couples to the average density inside the unit cell $\bar \rho(\vec x)$. On the other hand, the lattice-scale oscillation of bare $\rho(\vec x)$ renders it incompatible with continuum limit in long wavelength. $\rho(\vec x)$ for the boundary charge density, however,matches with $\vec{\tilde P}$(see last paragraph).
 
 Similarly, we get boundary $k_F$ as one varies chemical potential. The bulk polarization is calculated by Eq.~\eqref{PBerry}. For simplicity we put the positive ions at sites with integral coordinates in units of Bravais lattice vectors, i.e. site $(0,0)$ and its equivalents by lattice translations. The ions don't contribute to polarization in this way; Eq \eqref{PBerry} gives the entire polarization then.
 
 As we vary the parameters $t,\epsilon$,  the system enters multiple physical regimes with a gapped bulk. For example, the system hosts non-chiral edge states in Fig.~\ref{square_model}b and when chemical potential stays inside the bulk gap, the edge density $\rho$, Fermi momentum $k_F$ and bulk polarization $P$ always obeys Eq.~\eqref{LuttingerViolation}. (We note it's important for the bulk to remain insulating with the chemical potential in between the gap.) A relatively trivial scenario in Fig.~\ref{square_model}d is in the absence of edge states, the edge density equals bulk polarization in line with previous knowledge of polarization. A tricky case is when the model has a nonzero Chern number $C$ shown in Fig.~\ref{square_model}c and edge density $\rho$, $k_F$, polarization will change upon a large gauge transform along the orthogonal direction, i.e. gauge-dependent. We find that under a fixed gauge, Eq.~\eqref{LuttingerViolation} still holds given appropriate recipe for bulk polarization Eq.~\eqref{PBerry}, i.e.,  $2\pi C$ discontinuity of Wilson loop phase $\theta_{k_1}=\int dk_2 \bra u_{\vec k}|i\partial_{k_2}|u_{\vec k}\ket$ occurs only at $k_1=0$. In all cases, the momenta of monopole on a torus geometry satisfy $(k_1,k_2)=2\pi(-P_2,P_1)$ obtained by the method in appendix \ref{translation_berry}.

Finally we remark that all calculations above apply also to the inter-cell part of polarization $\tilde P$, when we calculate the boundary charge density as $\rho_0=\int_{-\infty}^{na} dx_2 \int _{A_{bd}} dx_1\rho (x_1,x_2)\quad (n\in Z)$ instead of eq \eqref{bd_charge}. The relation between these two reads \cite{Bardarson_2016}
\begin{equation}
\label{rho_bd}
    \rho_{bd}=\rho_0-\frac{1}{a}\int_{na-a/2}^{na+a/2}dx_2\int _{A_{bd}} dx_1 \rho (x_1,x_2)x_2,
\end{equation}
where we take $x_0=na$ in eq \eqref{bd_charge} and use the neutrality condition. In passing we remark this rewriting makes explicit the equivalence between the window function method and the charge density as derived in Appendix~\ref{PPtilde}. There the charge density in $2$ spatial dimensions reads
\be
\rho(\vec x_0)=\int _{\Omega_{x_0}} d^2 x\rho(\vec x)-\frac{1}{a}\int_{\Omega_{x_0}} \vec  (\vec x-\vec x_0)\cdot\nabla_\Omega \rho(\vec x),
\ee
where $\Omega_{x_0}$ is the unit cell at $\vec x_0$ and $\nabla_\Omega$ denotes gradient of $\rho$ w.r.t its value at the same sublattice in neighboring unit cells. When there's a boundary to vacuum, the boundary charge $\sum_{x_0}\rho(\vec x_0)$ has two parts: the first term integrates to $\rho_0$ and second term integrates to give bulk electric dipole moment. Hence it agrees with eq~\eqref{rho_bd}.  

It is now clear that $\rho_0$ extracts the excess charge at the boundary. The boundaries we considered preserve complete unit cells in bulk. This extra charge accumulation part depends solely on the inter unit cell structure. Hence we use the gauge recipe eq \eqref{LandauGauge} to calculate monopole momentum related to $\tilde P$, whose  $A$ fields reside only on bonds between different unit cells and indeed they agree with Berry connection integral eq \eqref{PBerry} using periodic function $\tilde u_{\vec k}(\vec r)$.

\bibliography{Polarization}

\begin{thebibliography}{53}%
\makeatletter
\providecommand \@ifxundefined [1]{%
 \@ifx{#1\undefined}
}%
\providecommand \@ifnum [1]{%
 \ifnum #1\expandafter \@firstoftwo
 \else \expandafter \@secondoftwo
 \fi
}%
\providecommand \@ifx [1]{%
 \ifx #1\expandafter \@firstoftwo
 \else \expandafter \@secondoftwo
 \fi
}%
\providecommand \natexlab [1]{#1}%
\providecommand \enquote  [1]{``#1''}%
\providecommand \bibnamefont  [1]{#1}%
\providecommand \bibfnamefont [1]{#1}%
\providecommand \citenamefont [1]{#1}%
\providecommand \href@noop [0]{\@secondoftwo}%
\providecommand \href [0]{\begingroup \@sanitize@url \@href}%
\providecommand \@href[1]{\@@startlink{#1}\@@href}%
\providecommand \@@href[1]{\endgroup#1\@@endlink}%
\providecommand \@sanitize@url [0]{\catcode `\\12\catcode `\$12\catcode
  `\&12\catcode `\#12\catcode `\^12\catcode `\_12\catcode `\%12\relax}%
\providecommand \@@startlink[1]{}%
\providecommand \@@endlink[0]{}%
\providecommand \url  [0]{\begingroup\@sanitize@url \@url }%
\providecommand \@url [1]{\endgroup\@href {#1}{\urlprefix }}%
\providecommand \urlprefix  [0]{URL }%
\providecommand \Eprint [0]{\href }%
\providecommand \doibase [0]{http://dx.doi.org/}%
\providecommand \selectlanguage [0]{\@gobble}%
\providecommand \bibinfo  [0]{\@secondoftwo}%
\providecommand \bibfield  [0]{\@secondoftwo}%
\providecommand \translation [1]{[#1]}%
\providecommand \BibitemOpen [0]{}%
\providecommand \bibitemStop [0]{}%
\providecommand \bibitemNoStop [0]{.\EOS\space}%
\providecommand \EOS [0]{\spacefactor3000\relax}%
\providecommand \BibitemShut  [1]{\csname bibitem#1\endcsname}%
\let\auto@bib@innerbib\@empty
%</preamble>
\bibitem [{\citenamefont {Vanderbilt}(2018)}]{vanderbiltbook}%
  \BibitemOpen
  \bibfield  {author} {\bibinfo {author} {\bibfnamefont {David}\ \bibnamefont
  {Vanderbilt}},\ }\bibfield  {title} {\enquote {\bibinfo {title} {Berry phases
  in electronic structure theory: Electric polarization, orbital magnetization
  and topological insulators},}\ }\href {\doibase 10.1017/9781316662205}
  {\bibfield  {journal} {\bibinfo  {journal} {Cambridge University Press}\ }
  (\bibinfo {year} {2018}),\ 10.1017/9781316662205}\BibitemShut {NoStop}%
\bibitem [{\citenamefont {Resta}(1994)}]{Resta_rmp}%
  \BibitemOpen
  \bibfield  {author} {\bibinfo {author} {\bibfnamefont {Raffaele}\
  \bibnamefont {Resta}},\ }\bibfield  {title} {\enquote {\bibinfo {title}
  {Macroscopic polarization in crystalline dielectrics: the geometric phase
  approach},}\ }\href {\doibase 10.1103/RevModPhys.66.899} {\bibfield
  {journal} {\bibinfo  {journal} {Rev. Mod. Phys.}\ }\textbf {\bibinfo {volume}
  {66}},\ \bibinfo {pages} {899--915} (\bibinfo {year} {1994})}\BibitemShut
  {NoStop}%
\bibitem [{\citenamefont {Resta}\ and\ \citenamefont
  {Vanderbilt}(2007)}]{RestaVanderbilt}%
  \BibitemOpen
  \bibfield  {author} {\bibinfo {author} {\bibfnamefont {Raffaele}\
  \bibnamefont {Resta}}\ and\ \bibinfo {author} {\bibfnamefont {David}\
  \bibnamefont {Vanderbilt}},\ }\enquote {\bibinfo {title} {Theory of
  polarization: A modern approach},}\ in\ \href {\doibase
  10.1007/978-3-540-34591-6_2} {\emph {\bibinfo {booktitle} {Physics of
  Ferroelectrics: A Modern Perspective}}}\ (\bibinfo  {publisher} {Springer
  Berlin Heidelberg},\ \bibinfo {address} {Berlin, Heidelberg},\ \bibinfo
  {year} {2007})\ pp.\ \bibinfo {pages} {31--68}\BibitemShut {NoStop}%
\bibitem [{\citenamefont {Martin}(1972)}]{Martin72}%
  \BibitemOpen
  \bibfield  {author} {\bibinfo {author} {\bibfnamefont {Richard~M.}\
  \bibnamefont {Martin}},\ }\bibfield  {title} {\enquote {\bibinfo {title}
  {Piezoelectricity},}\ }\href {\doibase 10.1103/PhysRevB.5.1607} {\bibfield
  {journal} {\bibinfo  {journal} {Phys. Rev. B}\ }\textbf {\bibinfo {volume}
  {5}},\ \bibinfo {pages} {1607--1613} (\bibinfo {year} {1972})}\BibitemShut
  {NoStop}%
\bibitem [{\citenamefont {Kallin}\ and\ \citenamefont
  {Halperin}(1984)}]{Kallin84}%
  \BibitemOpen
  \bibfield  {author} {\bibinfo {author} {\bibfnamefont {C.}~\bibnamefont
  {Kallin}}\ and\ \bibinfo {author} {\bibfnamefont {B.~I.}\ \bibnamefont
  {Halperin}},\ }\bibfield  {title} {\enquote {\bibinfo {title}
  {Surface-induced charge disturbances and piezoelectricity in insulating
  crystals},}\ }\href {\doibase 10.1103/PhysRevB.29.2175} {\bibfield  {journal}
  {\bibinfo  {journal} {Phys. Rev. B}\ }\textbf {\bibinfo {volume} {29}},\
  \bibinfo {pages} {2175--2189} (\bibinfo {year} {1984})}\BibitemShut {NoStop}%
\bibitem [{\citenamefont {King-Smith}\ and\ \citenamefont
  {Vanderbilt}(1993)}]{Vanderbilt}%
  \BibitemOpen
  \bibfield  {author} {\bibinfo {author} {\bibfnamefont {R.~D.}\ \bibnamefont
  {King-Smith}}\ and\ \bibinfo {author} {\bibfnamefont {David}\ \bibnamefont
  {Vanderbilt}},\ }\bibfield  {title} {\enquote {\bibinfo {title} {Theory of
  polarization of crystalline solids},}\ }\href {\doibase
  10.1103/PhysRevB.47.1651} {\bibfield  {journal} {\bibinfo  {journal} {Phys.
  Rev. B}\ }\textbf {\bibinfo {volume} {47}},\ \bibinfo {pages} {1651--1654}
  (\bibinfo {year} {1993})}\BibitemShut {NoStop}%
\bibitem [{\citenamefont {Resta}\ and\ \citenamefont
  {Sorella}(1995)}]{Resta95}%
  \BibitemOpen
  \bibfield  {author} {\bibinfo {author} {\bibfnamefont {R.}~\bibnamefont
  {Resta}}\ and\ \bibinfo {author} {\bibfnamefont {S.}~\bibnamefont
  {Sorella}},\ }\bibfield  {title} {\enquote {\bibinfo {title} {Many-body
  effects on polarization and dynamical charges in a partly covalent polar
  insulator},}\ }\href {\doibase 10.1103/PhysRevLett.74.4738} {\bibfield
  {journal} {\bibinfo  {journal} {Phys. Rev. Lett.}\ }\textbf {\bibinfo
  {volume} {74}},\ \bibinfo {pages} {4738--4741} (\bibinfo {year}
  {1995})}\BibitemShut {NoStop}%
\bibitem [{\citenamefont {Ortiz}\ \emph {et~al.}(1996)\citenamefont {Ortiz},
  \citenamefont {Ordej\'on}, \citenamefont {Martin},\ and\ \citenamefont
  {Chiappe}}]{Ortiz96}%
  \BibitemOpen
  \bibfield  {author} {\bibinfo {author} {\bibfnamefont {Gerardo}\ \bibnamefont
  {Ortiz}}, \bibinfo {author} {\bibfnamefont {Pablo}\ \bibnamefont
  {Ordej\'on}}, \bibinfo {author} {\bibfnamefont {Richard~M.}\ \bibnamefont
  {Martin}}, \ and\ \bibinfo {author} {\bibfnamefont {Guillermo}\ \bibnamefont
  {Chiappe}},\ }\bibfield  {title} {\enquote {\bibinfo {title} {Quantum phase
  transitions involving a change in polarization},}\ }\href {\doibase
  10.1103/PhysRevB.54.13515} {\bibfield  {journal} {\bibinfo  {journal} {Phys.
  Rev. B}\ }\textbf {\bibinfo {volume} {54}},\ \bibinfo {pages} {13515--13528}
  (\bibinfo {year} {1996})}\BibitemShut {NoStop}%
\bibitem [{\citenamefont {Martin}\ and\ \citenamefont
  {Ortiz}(1997)}]{Martin97}%
  \BibitemOpen
  \bibfield  {author} {\bibinfo {author} {\bibfnamefont {Richard~M.}\
  \bibnamefont {Martin}}\ and\ \bibinfo {author} {\bibfnamefont {Gerardo}\
  \bibnamefont {Ortiz}},\ }\bibfield  {title} {\enquote {\bibinfo {title}
  {Recent developments in the theory of electric polarization in solids},}\
  }\href {\doibase https://doi.org/10.1016/S0038-1098(96)00719-3} {\bibfield
  {journal} {\bibinfo  {journal} {Solid State Communications}\ }\textbf
  {\bibinfo {volume} {102}},\ \bibinfo {pages} {121 -- 126} (\bibinfo {year}
  {1997})},\ \bibinfo {note} {highlights in Condensed Matter Physics and
  Materials Science}\BibitemShut {NoStop}%
\bibitem [{\citenamefont {Martin}(1974)}]{Martin98}%
  \BibitemOpen
  \bibfield  {author} {\bibinfo {author} {\bibfnamefont {Richard~M.}\
  \bibnamefont {Martin}},\ }\bibfield  {title} {\enquote {\bibinfo {title}
  {Comment on calculations of electric polarization in crystals},}\ }\href
  {\doibase 10.1103/PhysRevB.9.1998} {\bibfield  {journal} {\bibinfo  {journal}
  {Phys. Rev. B}\ }\textbf {\bibinfo {volume} {9}},\ \bibinfo {pages}
  {1998--1999} (\bibinfo {year} {1974})}\BibitemShut {NoStop}%
\bibitem [{\citenamefont {{Watanabe}}\ and\ \citenamefont
  {{Oshikawa}}(2018)}]{WatanabeOshikawa}%
  \BibitemOpen
  \bibfield  {author} {\bibinfo {author} {\bibfnamefont {Haruki}\ \bibnamefont
  {{Watanabe}}}\ and\ \bibinfo {author} {\bibfnamefont {Masaki}\ \bibnamefont
  {{Oshikawa}}},\ }\bibfield  {title} {\enquote {\bibinfo {title}
  {{Inequivalent Berry Phases for the Bulk Polarization}},}\ }\href {\doibase
  10.1103/PhysRevX.8.021065} {\bibfield  {journal} {\bibinfo  {journal}
  {Physical Review X}\ }\textbf {\bibinfo {volume} {8}},\ \bibinfo {eid}
  {021065} (\bibinfo {year} {2018})},\ \Eprint
  {http://arxiv.org/abs/1802.00218} {arXiv:1802.00218 [cond-mat.str-el]}
  \BibitemShut {NoStop}%
\bibitem [{\citenamefont {Resta}(1992)}]{Resta92}%
  \BibitemOpen
  \bibfield  {author} {\bibinfo {author} {\bibfnamefont {R.}~\bibnamefont
  {Resta}},\ }\bibfield  {title} {\enquote {\bibinfo {title} {Theory of the
  electric polarization in crystals},}\ }\href {\doibase
  10.1080/00150199208016065} {\bibfield  {journal} {\bibinfo  {journal}
  {Ferroelectrics}\ }\textbf {\bibinfo {volume} {136}},\ \bibinfo {pages}
  {51--55} (\bibinfo {year} {1992})},\ \Eprint
  {http://arxiv.org/abs/https://doi.org/10.1080/00150199208016065}
  {https://doi.org/10.1080/00150199208016065} \BibitemShut {NoStop}%
\bibitem [{\citenamefont {Vanderbilt}\ and\ \citenamefont
  {King-Smith}(1993)}]{vanderbilt_1993}%
  \BibitemOpen
  \bibfield  {author} {\bibinfo {author} {\bibfnamefont {David}\ \bibnamefont
  {Vanderbilt}}\ and\ \bibinfo {author} {\bibfnamefont {R.~D.}\ \bibnamefont
  {King-Smith}},\ }\bibfield  {title} {\enquote {\bibinfo {title} {Electric
  polarization as a bulk quantity and its relation to surface charge},}\ }\href
  {\doibase 10.1103/PhysRevB.48.4442} {\bibfield  {journal} {\bibinfo
  {journal} {Phys. Rev. B}\ }\textbf {\bibinfo {volume} {48}},\ \bibinfo
  {pages} {4442--4455} (\bibinfo {year} {1993})}\BibitemShut {NoStop}%
\bibitem [{\citenamefont {Ortiz}\ and\ \citenamefont
  {Martin}(1994)}]{OrtizMartin}%
  \BibitemOpen
  \bibfield  {author} {\bibinfo {author} {\bibfnamefont {Gerardo}\ \bibnamefont
  {Ortiz}}\ and\ \bibinfo {author} {\bibfnamefont {Richard~M.}\ \bibnamefont
  {Martin}},\ }\bibfield  {title} {\enquote {\bibinfo {title} {Macroscopic
  polarization as a geometric quantum phase: Many-body formulation},}\ }\href
  {\doibase 10.1103/PhysRevB.49.14202} {\bibfield  {journal} {\bibinfo
  {journal} {Phys. Rev. B}\ }\textbf {\bibinfo {volume} {49}},\ \bibinfo
  {pages} {14202--14210} (\bibinfo {year} {1994})}\BibitemShut {NoStop}%
\bibitem [{\citenamefont {Kudin}\ \emph
  {et~al.}(2007{\natexlab{a}})\citenamefont {Kudin}, \citenamefont {Car},\ and\
  \citenamefont {Resta}}]{Kudin2}%
  \BibitemOpen
  \bibfield  {author} {\bibinfo {author} {\bibfnamefont {Konstantin~N.}\
  \bibnamefont {Kudin}}, \bibinfo {author} {\bibfnamefont {Roberto}\
  \bibnamefont {Car}}, \ and\ \bibinfo {author} {\bibfnamefont {Raffaele}\
  \bibnamefont {Resta}},\ }\bibfield  {title} {\enquote {\bibinfo {title}
  {Berry phase approach to longitudinal dipole moments of infinite chains in
  electronic-structure methods with local basis sets},}\ }\href {\doibase
  10.1063/1.2743018} {\bibfield  {journal} {\bibinfo  {journal} {The Journal of
  Chemical Physics}\ }\textbf {\bibinfo {volume} {126}},\ \bibinfo {pages}
  {234101} (\bibinfo {year} {2007}{\natexlab{a}})},\ \Eprint
  {http://arxiv.org/abs/https://doi.org/10.1063/1.2743018}
  {https://doi.org/10.1063/1.2743018} \BibitemShut {NoStop}%
\bibitem [{\citenamefont {Rhim}\ \emph {et~al.}(2017)\citenamefont {Rhim},
  \citenamefont {Behrends},\ and\ \citenamefont {Bardarson}}]{Bardarson_2016}%
  \BibitemOpen
  \bibfield  {author} {\bibinfo {author} {\bibfnamefont {Jun-Won}\ \bibnamefont
  {Rhim}}, \bibinfo {author} {\bibfnamefont {Jan}\ \bibnamefont {Behrends}}, \
  and\ \bibinfo {author} {\bibfnamefont {Jens~H.}\ \bibnamefont {Bardarson}},\
  }\bibfield  {title} {\enquote {\bibinfo {title} {Bulk-boundary correspondence
  from the intercellular zak phase},}\ }\href {\doibase
  10.1103/PhysRevB.95.035421} {\bibfield  {journal} {\bibinfo  {journal} {Phys.
  Rev. B}\ }\textbf {\bibinfo {volume} {95}},\ \bibinfo {pages} {035421}
  (\bibinfo {year} {2017})}\BibitemShut {NoStop}%
\bibitem [{\citenamefont {Resta}(1998{\natexlab{a}})}]{Resta}%
  \BibitemOpen
  \bibfield  {author} {\bibinfo {author} {\bibfnamefont {Raffaele}\
  \bibnamefont {Resta}},\ }\bibfield  {title} {\enquote {\bibinfo {title}
  {Quantum-mechanical position operator in extended systems},}\ }\href
  {\doibase 10.1103/PhysRevLett.80.1800} {\bibfield  {journal} {\bibinfo
  {journal} {Phys. Rev. Lett.}\ }\textbf {\bibinfo {volume} {80}},\ \bibinfo
  {pages} {1800--1803} (\bibinfo {year} {1998}{\natexlab{a}})}\BibitemShut
  {NoStop}%
\bibitem [{\citenamefont {Resta}(1998{\natexlab{b}})}]{resta_98}%
  \BibitemOpen
  \bibfield  {author} {\bibinfo {author} {\bibfnamefont {Raffaele}\
  \bibnamefont {Resta}},\ }\bibfield  {title} {\enquote {\bibinfo {title}
  {Quantum-mechanical position operator in extended systems},}\ }\href
  {\doibase 10.1103/PhysRevLett.80.1800} {\bibfield  {journal} {\bibinfo
  {journal} {Phys. Rev. Lett.}\ }\textbf {\bibinfo {volume} {80}},\ \bibinfo
  {pages} {1800--1803} (\bibinfo {year} {1998}{\natexlab{b}})}\BibitemShut
  {NoStop}%
\bibitem [{\citenamefont {{Aligia}}\ and\ \citenamefont
  {{Ortiz}}(1999)}]{aligia_99}%
  \BibitemOpen
  \bibfield  {author} {\bibinfo {author} {\bibfnamefont {A.~A.}\ \bibnamefont
  {{Aligia}}}\ and\ \bibinfo {author} {\bibfnamefont {G.}~\bibnamefont
  {{Ortiz}}},\ }\bibfield  {title} {\enquote {\bibinfo {title} {{Quantum
  Mechanical Position Operator and Localization in Extended Systems}},}\ }\href
  {\doibase 10.1103/PhysRevLett.82.2560} {\bibfield  {journal} {\bibinfo
  {journal} {\prl}\ }\textbf {\bibinfo {volume} {82}},\ \bibinfo {pages}
  {2560--2563} (\bibinfo {year} {1999})},\ \Eprint
  {http://arxiv.org/abs/cond-mat/9810348} {arXiv:cond-mat/9810348 [cond-mat]}
  \BibitemShut {NoStop}%
\bibitem [{\citenamefont {Thorngren}\ and\ \citenamefont
  {Else}(2018)}]{ThorngrenElse}%
  \BibitemOpen
  \bibfield  {author} {\bibinfo {author} {\bibfnamefont {Ryan}\ \bibnamefont
  {Thorngren}}\ and\ \bibinfo {author} {\bibfnamefont {Dominic~V.}\
  \bibnamefont {Else}},\ }\bibfield  {title} {\enquote {\bibinfo {title}
  {Gauging spatial symmetries and the classification of topological crystalline
  phases},}\ }\href {\doibase 10.1103/PhysRevX.8.011040} {\bibfield  {journal}
  {\bibinfo  {journal} {Phys. Rev. X}\ }\textbf {\bibinfo {volume} {8}},\
  \bibinfo {pages} {011040} (\bibinfo {year} {2018})}\BibitemShut {NoStop}%
\bibitem [{\citenamefont {Dzyaloshinskii}\ and\ \citenamefont
  {Volovick}(1980)}]{DZYALOSHINSKII198067}%
  \BibitemOpen
  \bibfield  {author} {\bibinfo {author} {\bibfnamefont {I.E.}\ \bibnamefont
  {Dzyaloshinskii}}\ and\ \bibinfo {author} {\bibfnamefont {G.E.}\ \bibnamefont
  {Volovick}},\ }\bibfield  {title} {\enquote {\bibinfo {title} {Poisson
  brackets in condensed matter physics},}\ }\href {\doibase
  https://doi.org/10.1016/0003-4916(80)90119-0} {\bibfield  {journal} {\bibinfo
   {journal} {Annals of Physics}\ }\textbf {\bibinfo {volume} {125}},\ \bibinfo
  {pages} {67 -- 97} (\bibinfo {year} {1980})}\BibitemShut {NoStop}%
\bibitem [{\citenamefont {Andreev}\ and\ \citenamefont
  {Kagan}(1984)}]{Tetrads1}%
  \BibitemOpen
  \bibfield  {author} {\bibinfo {author} {\bibfnamefont {A.F}\ \bibnamefont
  {Andreev}}\ and\ \bibinfo {author} {\bibfnamefont {M.Yu}\ \bibnamefont
  {Kagan}},\ }\bibfield  {title} {\enquote {\bibinfo {title} {Hydrodynamcs of a
  rotating superfluid liquid},}\ }\href@noop {} {\bibfield  {journal} {\bibinfo
   {journal} {Zh. Eksp. Teor. Fiz.}\ }\textbf {\bibinfo {volume} {86}},\
  \bibinfo {pages} {546--557} (\bibinfo {year} {1984})}\BibitemShut {NoStop}%
\bibitem [{\citenamefont {Nissinen}\ and\ \citenamefont
  {Volovik}(2018)}]{Tetrads2}%
  \BibitemOpen
  \bibfield  {author} {\bibinfo {author} {\bibfnamefont {J.}~\bibnamefont
  {Nissinen}}\ and\ \bibinfo {author} {\bibfnamefont {G.~E.}\ \bibnamefont
  {Volovik}},\ }\bibfield  {title} {\enquote {\bibinfo {title} {Tetrads in
  solids: from elasticity theory to topological quantum hall systems and weyl
  fermions},}\ }\href {\doibase 10.1134/S1063776118110080} {\bibfield
  {journal} {\bibinfo  {journal} {Journal of Experimental and Theoretical
  Physics}\ }\textbf {\bibinfo {volume} {127}},\ \bibinfo {pages} {948--957}
  (\bibinfo {year} {2018})}\BibitemShut {NoStop}%
\bibitem [{\citenamefont {Nissinen}\ and\ \citenamefont
  {Volovik}(2019)}]{Tetrads}%
  \BibitemOpen
  \bibfield  {author} {\bibinfo {author} {\bibfnamefont {J.}~\bibnamefont
  {Nissinen}}\ and\ \bibinfo {author} {\bibfnamefont {G.~E.}\ \bibnamefont
  {Volovik}},\ }\bibfield  {title} {\enquote {\bibinfo {title} {Elasticity
  tetrads, mixed axial-gravitational anomalies, and ($3+1$)-d quantum hall
  effect},}\ }\href {\doibase 10.1103/PhysRevResearch.1.023007} {\bibfield
  {journal} {\bibinfo  {journal} {Phys. Rev. Research}\ }\textbf {\bibinfo
  {volume} {1}},\ \bibinfo {pages} {023007} (\bibinfo {year}
  {2019})}\BibitemShut {NoStop}%
\bibitem [{\citenamefont {Huang}\ \emph {et~al.}(2019)\citenamefont {Huang},
  \citenamefont {Li}, \citenamefont {Zhou},\ and\ \citenamefont
  {Zhang}}]{torsion_huang}%
  \BibitemOpen
  \bibfield  {author} {\bibinfo {author} {\bibfnamefont {Ze-Min}\ \bibnamefont
  {Huang}}, \bibinfo {author} {\bibfnamefont {Longyue}\ \bibnamefont {Li}},
  \bibinfo {author} {\bibfnamefont {Jianhui}\ \bibnamefont {Zhou}}, \ and\
  \bibinfo {author} {\bibfnamefont {Hong-Hao}\ \bibnamefont {Zhang}},\
  }\bibfield  {title} {\enquote {\bibinfo {title} {Torsional response and
  liouville anomaly in weyl semimetals with dislocations},}\ }\href {\doibase
  10.1103/PhysRevB.99.155152} {\bibfield  {journal} {\bibinfo  {journal} {Phys.
  Rev. B}\ }\textbf {\bibinfo {volume} {99}},\ \bibinfo {pages} {155152}
  (\bibinfo {year} {2019})}\BibitemShut {NoStop}%
\bibitem [{\citenamefont {Kleinert}(1989)}]{kleinert}%
  \BibitemOpen
  \bibfield  {author} {\bibinfo {author} {\bibfnamefont {H}~\bibnamefont
  {Kleinert}},\ }\href {\doibase 10.1142/0356} {\emph {\bibinfo {title} {Gauge
  Fields in Condensed Matter}}}\ (\bibinfo  {publisher} {WORLD SCIENTIFIC},\
  \bibinfo {year} {1989})\ \Eprint
  {http://arxiv.org/abs/https://www.worldscientific.com/doi/pdf/10.1142/0356}
  {https://www.worldscientific.com/doi/pdf/10.1142/0356} \BibitemShut {NoStop}%
\bibitem [{\citenamefont {{Manjunath}}\ and\ \citenamefont
  {{Barkeshli}}(2020)}]{Manjunath_2020}%
  \BibitemOpen
  \bibfield  {author} {\bibinfo {author} {\bibfnamefont {Naren}\ \bibnamefont
  {{Manjunath}}}\ and\ \bibinfo {author} {\bibfnamefont {Maissam}\ \bibnamefont
  {{Barkeshli}}},\ }\bibfield  {title} {\enquote {\bibinfo {title}
  {{Crystalline gauge fields and quantized discrete geometric response for
  Abelian topological phases with lattice symmetry}},}\ }\href@noop {}
  {\bibfield  {journal} {\bibinfo  {journal} {arXiv e-prints}\ ,\ \bibinfo
  {eid} {arXiv:2005.10265}} (\bibinfo {year} {2020})},\ \Eprint
  {http://arxiv.org/abs/2005.10265} {arXiv:2005.10265 [cond-mat.str-el]}
  \BibitemShut {NoStop}%
\bibitem [{Note1()}]{Note1}%
  \BibitemOpen
  \bibinfo {note} {This integrating by part may leave a boundary term which
  accounts for the change of surface bound charge $\sigma _i$ due to spatial
  variation of polarization through $\sigma _i={\protect \bf P}\cdot {\setbox
  \z@ \hbox {\frozen@everymath \@emptytoks \mathsurround \z@ $\textstyle
  n$}\mathaccent "0362{n}}$, where ${\setbox \z@ \hbox {\frozen@everymath
  \@emptytoks \mathsurround \z@ $\textstyle n$}\mathaccent "0362{n}}$ is the
  normal vector of the surface.}\BibitemShut {Stop}%
\bibitem [{\citenamefont {Dijkgraaf}\ and\ \citenamefont
  {Witten}(1990)}]{DijkgraafWitten}%
  \BibitemOpen
  \bibfield  {author} {\bibinfo {author} {\bibfnamefont {Robbert}\ \bibnamefont
  {Dijkgraaf}}\ and\ \bibinfo {author} {\bibfnamefont {Edward}\ \bibnamefont
  {Witten}},\ }\bibfield  {title} {\enquote {\bibinfo {title} {Topological
  gauge theories and group cohomology},}\ }\href {\doibase 10.1007/BF02096988}
  {\bibfield  {journal} {\bibinfo  {journal} {Communications in Mathematical
  Physics}\ }\textbf {\bibinfo {volume} {129}},\ \bibinfo {pages} {393--429}
  (\bibinfo {year} {1990})}\BibitemShut {NoStop}%
\bibitem [{\citenamefont {Alicea}(2008)}]{alicea_2008}%
  \BibitemOpen
  \bibfield  {author} {\bibinfo {author} {\bibfnamefont {Jason}\ \bibnamefont
  {Alicea}},\ }\bibfield  {title} {\enquote {\bibinfo {title} {Monopole quantum
  numbers in the staggered flux spin liquid},}\ }\href {\doibase
  10.1103/PhysRevB.78.035126} {\bibfield  {journal} {\bibinfo  {journal} {Phys.
  Rev. B}\ }\textbf {\bibinfo {volume} {78}},\ \bibinfo {pages} {035126}
  (\bibinfo {year} {2008})}\BibitemShut {NoStop}%
\bibitem [{\citenamefont {{Ran}}\ \emph {et~al.}(2008)\citenamefont {{Ran}},
  \citenamefont {{Vishwanath}},\ and\ \citenamefont
  {{Lee}}}]{ranvishwanathlee}%
  \BibitemOpen
  \bibfield  {author} {\bibinfo {author} {\bibfnamefont {Y.}~\bibnamefont
  {{Ran}}}, \bibinfo {author} {\bibfnamefont {A.}~\bibnamefont {{Vishwanath}}},
  \ and\ \bibinfo {author} {\bibfnamefont {D.-H.}\ \bibnamefont {{Lee}}},\
  }\bibfield  {title} {\enquote {\bibinfo {title} {{A direct transition between
  a Neel ordered Mott insulator and a $d\_{x^2-y^2}$ superconductor on the
  square lattice}},}\ }\href@noop {} {\bibfield  {journal} {\bibinfo  {journal}
  {ArXiv e-prints}\ } (\bibinfo {year} {2008})},\ \Eprint
  {http://arxiv.org/abs/0806.2321} {arXiv:0806.2321 [cond-mat.str-el]}
  \BibitemShut {NoStop}%
\bibitem [{\citenamefont {Hermele}\ \emph {et~al.}(2008)\citenamefont
  {Hermele}, \citenamefont {Ran}, \citenamefont {Lee},\ and\ \citenamefont
  {Wen}}]{hermele_2008}%
  \BibitemOpen
  \bibfield  {author} {\bibinfo {author} {\bibfnamefont {Michael}\ \bibnamefont
  {Hermele}}, \bibinfo {author} {\bibfnamefont {Ying}\ \bibnamefont {Ran}},
  \bibinfo {author} {\bibfnamefont {Patrick~A.}\ \bibnamefont {Lee}}, \ and\
  \bibinfo {author} {\bibfnamefont {Xiao-Gang}\ \bibnamefont {Wen}},\
  }\bibfield  {title} {\enquote {\bibinfo {title} {Properties of an algebraic
  spin liquid on the kagome lattice},}\ }\href {\doibase
  10.1103/PhysRevB.77.224413} {\bibfield  {journal} {\bibinfo  {journal} {Phys.
  Rev. B}\ }\textbf {\bibinfo {volume} {77}},\ \bibinfo {pages} {224413}
  (\bibinfo {year} {2008})}\BibitemShut {NoStop}%
\bibitem [{\citenamefont {Song}\ \emph {et~al.}(2019)\citenamefont {Song},
  \citenamefont {Wang}, \citenamefont {Vishwanath},\ and\ \citenamefont
  {He}}]{monopole2}%
  \BibitemOpen
  \bibfield  {author} {\bibinfo {author} {\bibfnamefont {Xue-Yang}\
  \bibnamefont {Song}}, \bibinfo {author} {\bibfnamefont {Chong}\ \bibnamefont
  {Wang}}, \bibinfo {author} {\bibfnamefont {Ashvin}\ \bibnamefont
  {Vishwanath}}, \ and\ \bibinfo {author} {\bibfnamefont {Yin-Chen}\
  \bibnamefont {He}},\ }\bibfield  {title} {\enquote {\bibinfo {title}
  {Unifying description of competing orders in two-dimensional quantum
  magnets},}\ }\href {\doibase 10.1038/s41467-019-11727-3} {\bibfield
  {journal} {\bibinfo  {journal} {Nature Communications}\ }\textbf {\bibinfo
  {volume} {10}},\ \bibinfo {pages} {4254} (\bibinfo {year}
  {2019})}\BibitemShut {NoStop}%
\bibitem [{\citenamefont {Pollmann}\ \emph {et~al.}(2010)\citenamefont
  {Pollmann}, \citenamefont {Turner}, \citenamefont {Berg},\ and\ \citenamefont
  {Oshikawa}}]{Pollmann2010}%
  \BibitemOpen
  \bibfield  {author} {\bibinfo {author} {\bibfnamefont {Frank}\ \bibnamefont
  {Pollmann}}, \bibinfo {author} {\bibfnamefont {Ari~M.}\ \bibnamefont
  {Turner}}, \bibinfo {author} {\bibfnamefont {Erez}\ \bibnamefont {Berg}}, \
  and\ \bibinfo {author} {\bibfnamefont {Masaki}\ \bibnamefont {Oshikawa}},\
  }\bibfield  {title} {\enquote {\bibinfo {title} {Entanglement spectrum of a
  topological phase in one dimension},}\ }\href {\doibase
  10.1103/PhysRevB.81.064439} {\bibfield  {journal} {\bibinfo  {journal} {Phys.
  Rev. B}\ }\textbf {\bibinfo {volume} {81}},\ \bibinfo {pages} {064439}
  (\bibinfo {year} {2010})}\BibitemShut {NoStop}%
\bibitem [{\citenamefont {Chen}\ \emph {et~al.}(2011)\citenamefont {Chen},
  \citenamefont {Gu},\ and\ \citenamefont {Wen}}]{Chen2011}%
  \BibitemOpen
  \bibfield  {author} {\bibinfo {author} {\bibfnamefont {Xie}\ \bibnamefont
  {Chen}}, \bibinfo {author} {\bibfnamefont {Zheng-Cheng}\ \bibnamefont {Gu}},
  \ and\ \bibinfo {author} {\bibfnamefont {Xiao-Gang}\ \bibnamefont {Wen}},\
  }\bibfield  {title} {\enquote {\bibinfo {title} {Classification of gapped
  symmetric phases in one-dimensional spin systems},}\ }\href {\doibase
  10.1103/PhysRevB.83.035107} {\bibfield  {journal} {\bibinfo  {journal} {Phys.
  Rev. B}\ }\textbf {\bibinfo {volume} {83}},\ \bibinfo {pages} {035107}
  (\bibinfo {year} {2011})}\BibitemShut {NoStop}%
\bibitem [{\citenamefont {Schuch}\ \emph {et~al.}(2011)\citenamefont {Schuch},
  \citenamefont {P\'erez-Garc\'{\i}a},\ and\ \citenamefont
  {Cirac}}]{Schuch2011}%
  \BibitemOpen
  \bibfield  {author} {\bibinfo {author} {\bibfnamefont {Norbert}\ \bibnamefont
  {Schuch}}, \bibinfo {author} {\bibfnamefont {David}\ \bibnamefont
  {P\'erez-Garc\'{\i}a}}, \ and\ \bibinfo {author} {\bibfnamefont {Ignacio}\
  \bibnamefont {Cirac}},\ }\bibfield  {title} {\enquote {\bibinfo {title}
  {Classifying quantum phases using matrix product states and projected
  entangled pair states},}\ }\href {\doibase 10.1103/PhysRevB.84.165139}
  {\bibfield  {journal} {\bibinfo  {journal} {Phys. Rev. B}\ }\textbf {\bibinfo
  {volume} {84}},\ \bibinfo {pages} {165139} (\bibinfo {year}
  {2011})}\BibitemShut {NoStop}%
\bibitem [{\citenamefont {Savary}\ and\ \citenamefont
  {Balents}(2017)}]{savary_2017}%
  \BibitemOpen
  \bibfield  {author} {\bibinfo {author} {\bibfnamefont {Lucile}\ \bibnamefont
  {Savary}}\ and\ \bibinfo {author} {\bibfnamefont {Leon}\ \bibnamefont
  {Balents}},\ }\bibfield  {title} {\enquote {\bibinfo {title} {Quantum spin
  liquids: a review},}\ }\href
  {http://stacks.iop.org/0034-4885/80/i=1/a=016502} {\bibfield  {journal}
  {\bibinfo  {journal} {Reports on Progress in Physics}\ }\textbf {\bibinfo
  {volume} {80}},\ \bibinfo {pages} {016502} (\bibinfo {year}
  {2017})}\BibitemShut {NoStop}%
\bibitem [{\citenamefont {Oshikawa}(2000)}]{Oshikawa}%
  \BibitemOpen
  \bibfield  {author} {\bibinfo {author} {\bibfnamefont {Masaki}\ \bibnamefont
  {Oshikawa}},\ }\bibfield  {title} {\enquote {\bibinfo {title} {Topological
  approach to luttinger's theorem and the fermi surface of a kondo lattice},}\
  }\href {\doibase 10.1103/PhysRevLett.84.3370} {\bibfield  {journal} {\bibinfo
   {journal} {Phys. Rev. Lett.}\ }\textbf {\bibinfo {volume} {84}},\ \bibinfo
  {pages} {3370--3373} (\bibinfo {year} {2000})}\BibitemShut {NoStop}%
\bibitem [{\citenamefont {Lieb}\ \emph {et~al.}(1961)\citenamefont {Lieb},
  \citenamefont {Schultz},\ and\ \citenamefont {Mattis}}]{LSM}%
  \BibitemOpen
  \bibfield  {author} {\bibinfo {author} {\bibfnamefont {Elliott}\ \bibnamefont
  {Lieb}}, \bibinfo {author} {\bibfnamefont {Theodore}\ \bibnamefont
  {Schultz}}, \ and\ \bibinfo {author} {\bibfnamefont {Daniel}\ \bibnamefont
  {Mattis}},\ }\bibfield  {title} {\enquote {\bibinfo {title} {Two soluble
  models of an antiferromagnetic chain},}\ }\href {\doibase
  https://doi.org/10.1016/0003-4916(61)90115-4} {\bibfield  {journal} {\bibinfo
   {journal} {Annals of Physics}\ }\textbf {\bibinfo {volume} {16}},\ \bibinfo
  {pages} {407 -- 466} (\bibinfo {year} {1961})}\BibitemShut {NoStop}%
\bibitem [{\citenamefont {Hastings}(2004)}]{Hasting}%
  \BibitemOpen
  \bibfield  {author} {\bibinfo {author} {\bibfnamefont {M.~B.}\ \bibnamefont
  {Hastings}},\ }\bibfield  {title} {\enquote {\bibinfo {title}
  {Lieb-schultz-mattis in higher dimensions},}\ }\href {\doibase
  10.1103/PhysRevB.69.104431} {\bibfield  {journal} {\bibinfo  {journal} {Phys.
  Rev. B}\ }\textbf {\bibinfo {volume} {69}},\ \bibinfo {pages} {104431}
  (\bibinfo {year} {2004})}\BibitemShut {NoStop}%
\bibitem [{\citenamefont {Cheng}\ \emph {et~al.}(2016)\citenamefont {Cheng},
  \citenamefont {Zaletel}, \citenamefont {Barkeshli}, \citenamefont
  {Vishwanath},\ and\ \citenamefont {Bonderson}}]{Cheng_2016}%
  \BibitemOpen
  \bibfield  {author} {\bibinfo {author} {\bibfnamefont {Meng}\ \bibnamefont
  {Cheng}}, \bibinfo {author} {\bibfnamefont {Michael}\ \bibnamefont
  {Zaletel}}, \bibinfo {author} {\bibfnamefont {Maissam}\ \bibnamefont
  {Barkeshli}}, \bibinfo {author} {\bibfnamefont {Ashvin}\ \bibnamefont
  {Vishwanath}}, \ and\ \bibinfo {author} {\bibfnamefont {Parsa}\ \bibnamefont
  {Bonderson}},\ }\bibfield  {title} {\enquote {\bibinfo {title} {Translational
  symmetry and microscopic constraints on symmetry-enriched topological phases:
  A view from the surface},}\ }\href {\doibase 10.1103/PhysRevX.6.041068}
  {\bibfield  {journal} {\bibinfo  {journal} {Phys. Rev. X}\ }\textbf {\bibinfo
  {volume} {6}},\ \bibinfo {pages} {041068} (\bibinfo {year}
  {2016})}\BibitemShut {NoStop}%
\bibitem [{\citenamefont {{Metlitski}}\ and\ \citenamefont
  {{Thorngren}}(2018)}]{MetlitskiThorngren}%
  \BibitemOpen
  \bibfield  {author} {\bibinfo {author} {\bibfnamefont {M.~A.}\ \bibnamefont
  {{Metlitski}}}\ and\ \bibinfo {author} {\bibfnamefont {R.}~\bibnamefont
  {{Thorngren}}},\ }\bibfield  {title} {\enquote {\bibinfo {title} {{Intrinsic
  and emergent anomalies at deconfined critical points}},}\ }\href {\doibase
  10.1103/PhysRevB.98.085140} {\bibfield  {journal} {\bibinfo  {journal}
  {\prb}\ }\textbf {\bibinfo {volume} {98}},\ \bibinfo {eid} {085140} (\bibinfo
  {year} {2018})},\ \Eprint {http://arxiv.org/abs/1707.07686} {arXiv:1707.07686
  [cond-mat.str-el]} \BibitemShut {NoStop}%
\bibitem [{\citenamefont {Cho}\ \emph {et~al.}(2017)\citenamefont {Cho},
  \citenamefont {Hsieh},\ and\ \citenamefont {Ryu}}]{ChoHsiehRyu}%
  \BibitemOpen
  \bibfield  {author} {\bibinfo {author} {\bibfnamefont {Gil~Young}\
  \bibnamefont {Cho}}, \bibinfo {author} {\bibfnamefont {Chang-Tse}\
  \bibnamefont {Hsieh}}, \ and\ \bibinfo {author} {\bibfnamefont {Shinsei}\
  \bibnamefont {Ryu}},\ }\bibfield  {title} {\enquote {\bibinfo {title}
  {Anomaly manifestation of lieb-schultz-mattis theorem and topological
  phases},}\ }\href {\doibase 10.1103/PhysRevB.96.195105} {\bibfield  {journal}
  {\bibinfo  {journal} {Phys. Rev. B}\ }\textbf {\bibinfo {volume} {96}},\
  \bibinfo {pages} {195105} (\bibinfo {year} {2017})}\BibitemShut {NoStop}%
\bibitem [{\citenamefont {Teo}\ \emph {et~al.}(2008)\citenamefont {Teo},
  \citenamefont {Fu},\ and\ \citenamefont {Kane}}]{KaneLuttinger}%
  \BibitemOpen
  \bibfield  {author} {\bibinfo {author} {\bibfnamefont {Jeffrey C.~Y.}\
  \bibnamefont {Teo}}, \bibinfo {author} {\bibfnamefont {Liang}\ \bibnamefont
  {Fu}}, \ and\ \bibinfo {author} {\bibfnamefont {C.~L.}\ \bibnamefont
  {Kane}},\ }\bibfield  {title} {\enquote {\bibinfo {title} {Surface states and
  topological invariants in three-dimensional topological insulators:
  Application to ${\text{bi}}_{1\ensuremath{-}x}{\text{sb}}_{x}$},}\ }\href
  {\doibase 10.1103/PhysRevB.78.045426} {\bibfield  {journal} {\bibinfo
  {journal} {Phys. Rev. B}\ }\textbf {\bibinfo {volume} {78}},\ \bibinfo
  {pages} {045426} (\bibinfo {year} {2008})}\BibitemShut {NoStop}%
\bibitem [{\citenamefont {Coh}\ and\ \citenamefont
  {Vanderbilt}(2009)}]{chern_band}%
  \BibitemOpen
  \bibfield  {author} {\bibinfo {author} {\bibfnamefont {Sinisa}\ \bibnamefont
  {Coh}}\ and\ \bibinfo {author} {\bibfnamefont {David}\ \bibnamefont
  {Vanderbilt}},\ }\bibfield  {title} {\enquote {\bibinfo {title} {Electric
  polarization in a chern insulator},}\ }\href {\doibase
  10.1103/PhysRevLett.102.107603} {\bibfield  {journal} {\bibinfo  {journal}
  {Phys. Rev. Lett.}\ }\textbf {\bibinfo {volume} {102}},\ \bibinfo {pages}
  {107603} (\bibinfo {year} {2009})}\BibitemShut {NoStop}%
\bibitem [{\citenamefont {{van Miert}}\ and\ \citenamefont
  {{Ortix}}(2018)}]{DislocationCharge}%
  \BibitemOpen
  \bibfield  {author} {\bibinfo {author} {\bibfnamefont {Guido}\ \bibnamefont
  {{van Miert}}}\ and\ \bibinfo {author} {\bibfnamefont {Carmine}\ \bibnamefont
  {{Ortix}}},\ }\bibfield  {title} {\enquote {\bibinfo {title} {{Dislocation
  charges reveal two-dimensional topological crystalline invariants}},}\ }\href
  {\doibase 10.1103/PhysRevB.97.201111} {\bibfield  {journal} {\bibinfo
  {journal} {Physical Review B}\ }\textbf {\bibinfo {volume} {97}},\ \bibinfo
  {eid} {201111} (\bibinfo {year} {2018})},\ \Eprint
  {http://arxiv.org/abs/1802.00715} {arXiv:1802.00715 [cond-mat.mes-hall]}
  \BibitemShut {NoStop}%
\bibitem [{\citenamefont {{Song}}\ \emph {et~al.}(2018)\citenamefont {{Song}},
  \citenamefont {{He}}, \citenamefont {{Vishwanath}},\ and\ \citenamefont
  {{Wang}}}]{monopole1}%
  \BibitemOpen
  \bibfield  {author} {\bibinfo {author} {\bibfnamefont {Xue-Yang}\
  \bibnamefont {{Song}}}, \bibinfo {author} {\bibfnamefont {Yin-Chen}\
  \bibnamefont {{He}}}, \bibinfo {author} {\bibfnamefont {Ashvin}\ \bibnamefont
  {{Vishwanath}}}, \ and\ \bibinfo {author} {\bibfnamefont {Chong}\
  \bibnamefont {{Wang}}},\ }\bibfield  {title} {\enquote {\bibinfo {title}
  {{From spinon band topology to the symmetry quantum numbers of monopoles in
  Dirac spin liquids}},}\ }\href@noop {} {\bibfield  {journal} {\bibinfo
  {journal} {arXiv e-prints}\ ,\ \bibinfo {eid} {arXiv:1811.11182}} (\bibinfo
  {year} {2018})},\ \Eprint {http://arxiv.org/abs/1811.11182} {arXiv:1811.11182
  [cond-mat.str-el]} \BibitemShut {NoStop}%
\bibitem [{\citenamefont {{Qi}}\ \emph {et~al.}(2008)\citenamefont {{Qi}},
  \citenamefont {{Hughes}},\ and\ \citenamefont {{Zhang}}}]{QiHughesZhang}%
  \BibitemOpen
  \bibfield  {author} {\bibinfo {author} {\bibfnamefont {Xiao-Liang}\
  \bibnamefont {{Qi}}}, \bibinfo {author} {\bibfnamefont {Taylor~L.}\
  \bibnamefont {{Hughes}}}, \ and\ \bibinfo {author} {\bibfnamefont
  {Shou-Cheng}\ \bibnamefont {{Zhang}}},\ }\bibfield  {title} {\enquote
  {\bibinfo {title} {{Topological field theory of time-reversal invariant
  insulators}},}\ }\href {\doibase 10.1103/PhysRevB.78.195424} {\bibfield
  {journal} {\bibinfo  {journal} {\prb}\ }\textbf {\bibinfo {volume} {78}},\
  \bibinfo {eid} {195424} (\bibinfo {year} {2008})},\ \Eprint
  {http://arxiv.org/abs/0802.3537} {arXiv:0802.3537 [cond-mat.mes-hall]}
  \BibitemShut {NoStop}%
\bibitem [{\citenamefont {{Essin}}\ \emph {et~al.}(2008)\citenamefont
  {{Essin}}, \citenamefont {{Moore}},\ and\ \citenamefont
  {{Vanderbilt}}}]{EssinMooreVanderbilt}%
  \BibitemOpen
  \bibfield  {author} {\bibinfo {author} {\bibfnamefont {Andrew~M.}\
  \bibnamefont {{Essin}}}, \bibinfo {author} {\bibfnamefont {Joel~E.}\
  \bibnamefont {{Moore}}}, \ and\ \bibinfo {author} {\bibfnamefont {David}\
  \bibnamefont {{Vanderbilt}}},\ }\bibfield  {title} {\enquote {\bibinfo
  {title} {{Magnetoelectric polarizability and axion electrodynamics in
  crystalline insulators}},}\ }\href@noop {} {\bibfield  {journal} {\bibinfo
  {journal} {arXiv e-prints}\ ,\ \bibinfo {eid} {arXiv:0810.2998}} (\bibinfo
  {year} {2008})},\ \Eprint {http://arxiv.org/abs/0810.2998} {arXiv:0810.2998
  [cond-mat.mes-hall]} \BibitemShut {NoStop}%
\bibitem [{\citenamefont {{Witten}}(1979)}]{WittenDyon}%
  \BibitemOpen
  \bibfield  {author} {\bibinfo {author} {\bibfnamefont {E.}~\bibnamefont
  {{Witten}}},\ }\bibfield  {title} {\enquote {\bibinfo {title} {{Dyons of
  charge e{$\theta$}/2{$\pi$}}},}\ }\href {\doibase
  10.1016/0370-2693(79)90838-4} {\bibfield  {journal} {\bibinfo  {journal}
  {Physics Letters B}\ }\textbf {\bibinfo {volume} {86}},\ \bibinfo {pages}
  {283--287} (\bibinfo {year} {1979})}\BibitemShut {NoStop}%
\bibitem [{\citenamefont {{Lu}}\ \emph {et~al.}(2017)\citenamefont {{Lu}},
  \citenamefont {{Ran}},\ and\ \citenamefont {{Oshikawa}}}]{LuRanOshikawa}%
  \BibitemOpen
  \bibfield  {author} {\bibinfo {author} {\bibfnamefont {Yuan-Ming}\
  \bibnamefont {{Lu}}}, \bibinfo {author} {\bibfnamefont {Ying}\ \bibnamefont
  {{Ran}}}, \ and\ \bibinfo {author} {\bibfnamefont {Masaki}\ \bibnamefont
  {{Oshikawa}}},\ }\bibfield  {title} {\enquote {\bibinfo {title}
  {{Filling-enforced constraint on the quantized Hall conductivity on a
  periodic lattice}},}\ }\href@noop {} {\bibfield  {journal} {\bibinfo
  {journal} {arXiv e-prints}\ ,\ \bibinfo {eid} {arXiv:1705.09298}} (\bibinfo
  {year} {2017})},\ \Eprint {http://arxiv.org/abs/1705.09298} {arXiv:1705.09298
  [cond-mat.str-el]} \BibitemShut {NoStop}%
\bibitem [{\citenamefont {{Wang}}\ and\ \citenamefont
  {{Levin}}(2014)}]{WangLevin}%
  \BibitemOpen
  \bibfield  {author} {\bibinfo {author} {\bibfnamefont {Chenjie}\ \bibnamefont
  {{Wang}}}\ and\ \bibinfo {author} {\bibfnamefont {Michael}\ \bibnamefont
  {{Levin}}},\ }\bibfield  {title} {\enquote {\bibinfo {title} {{Braiding
  Statistics of Loop Excitations in Three Dimensions}},}\ }\href {\doibase
  10.1103/PhysRevLett.113.080403} {\bibfield  {journal} {\bibinfo  {journal}
  {\prl}\ }\textbf {\bibinfo {volume} {113}},\ \bibinfo {eid} {080403}
  (\bibinfo {year} {2014})},\ \Eprint {http://arxiv.org/abs/1403.7437}
  {arXiv:1403.7437 [cond-mat.str-el]} \BibitemShut {NoStop}%
\bibitem [{\citenamefont {Kudin}\ \emph
  {et~al.}(2007{\natexlab{b}})\citenamefont {Kudin}, \citenamefont {Car},\ and\
  \citenamefont {Resta}}]{Kudin_2007}%
  \BibitemOpen
  \bibfield  {author} {\bibinfo {author} {\bibfnamefont {Konstantin~N.}\
  \bibnamefont {Kudin}}, \bibinfo {author} {\bibfnamefont {Roberto}\
  \bibnamefont {Car}}, \ and\ \bibinfo {author} {\bibfnamefont {Raffaele}\
  \bibnamefont {Resta}},\ }\bibfield  {title} {\enquote {\bibinfo {title}
  {Quantization of the dipole moment and of the end charges in push-pull
  polymers},}\ }\href {\doibase 10.1063/1.2799514} {\bibfield  {journal}
  {\bibinfo  {journal} {The Journal of Chemical Physics}\ }\textbf {\bibinfo
  {volume} {127}},\ \bibinfo {pages} {194902} (\bibinfo {year}
  {2007}{\natexlab{b}})},\ \Eprint
  {http://arxiv.org/abs/https://doi.org/10.1063/1.2799514}
  {https://doi.org/10.1063/1.2799514} \BibitemShut {NoStop}%
\end{thebibliography}%

\end{document}